\pdfoutput=1

\RequirePackage{fix-cm}
\documentclass[twocolumn,epjc3]{svjour3}  
\usepackage{bbm}
\usepackage{graphicx}
\usepackage{slashed}
\usepackage{subfigure}
\usepackage[utf8]{inputenc}
\graphicspath{ {./images/} }
\usepackage{mathtools}

\newcommand{\comment}[1]{}
\newcommand{\cT}{\mathbf{T}}

\newcommand{\Tr}{\mathrm{Tr}}

\newcommand{\td}{\mathrm{d}}

\newcommand{\TT}[1]{\mathrm{#1}}

\newcommand{\qqquad}{\qquad\qquad}

\newcommand{\Nc}{N_{\mathrm{c}}}
\newcommand{\As}{\alpha_{\mathrm{s}}}
\usepackage{breakurl} 

\journalname{Eur. Phys. J. C}

\begin{document} 

\title{\boldmath Improvements on dipole shower colour}

\author{Jack Holguin \thanksref{e1,addr1} \and Jeffrey R. Forshaw \thanksref{e2,addr1}\and Simom Pl\"atzer \thanksref{e3,addr2}}
\thankstext{e1}{e-mail: jack.holguin@manchester.ac.uk}
\thankstext{e2}{e-mail: jeffrey.forshaw@manchester.ac.uk}
\thankstext{e3}{e-mail: simon.plaetzer@univie.ac.at}
\institute{Consortium for Fundamental Physics, School of Physics \& Astronomy, \\ University of Manchester, Manchester M13 9PL, United Kingdom \label{addr1} \and Particle Physics, Faculty of Physics, \\ University of Vienna, 1090 Wien, Austria \label{addr2}}

\date{\today}

\maketitle

\abstract{The dipole formalism provides a powerful framework from which parton showers can be constructed. In a recent paper \cite{Forshaw:2020wrq}, we proposed a dipole shower with improved colour accuracy and in this paper we show how it can be further improved. After an explicit check at $\mathcal{O}(\As^{2})$ we confirm that our original shower performs as it was designed to, i.e. inheriting its handling of angular-ordered radiation from a coherent branching algorithm. We also show how other dipole shower algorithms fail to achieve this. Nevertheless, there is an $\mathcal{O}(\As^{2})$ topology where it differs at sub-leading $\Nc$ from a coherent branching algorithm. This erroneous topology can contribute a leading logarithm to some observables and corresponds to emissions that are ordered in $k_t$ but not angle. We propose a simple, computationally efficient way to correct this and assign colour factors in accordance with the coherence properties of QCD to all orders in $\As$.}

\section{Introduction}
\label{sec:intro}

Parton showers typically are constructed using one of two basic approaches: angular-ordered showers (based on the coherent branching formalism) and dipole showers. Angular ordering is a very powerful approach, providing next-to-leading logarithmic (NLL) accuracy in some observables,\footnote{Many $e^+ e^-$ observables share the property that their distributions exponentiate: $$\Sigma(\As,L) = (1+C(\As))\exp(L \, g_{1}(\As L) + g_{2}(\As L) + ...),$$where $\Sigma$ is the fraction of events for which the observable is less than some value, $v = e^{-L}$. NLL accuracy corresponds to correctly computing the functions $g_{1}$ and $g_{2}$ \cite{Resum_large_logs_ee,Banfi:2004yd}.} but it fails to capture physics salient to the description of multi-jet final states in hadron colliders and non-global observables. By comparison, dipole showers are typically restricted to leading-colour accuracy but they can be applied across the board. In recent literature, much attention has been focused on improving the framework upon which dipole showers are constructed \cite{Pythia8,Gleisberg:2008ta,Herwig_dipole_shower,Hoche:2017iem,Hoche:2017hno,Cabouat:2017rzi,Dulat:2018vuy,Sjostrand:2019zhc,Cabouat:2019gtm,Cabouat:2020ssr,Buckley:2019kjt}. Substantial progress has been made demonstrating their capacity for NLL resummation \cite{Dasgupta:2020fwr,Forshaw:2020wrq} and methods for partially addressing sub-leading colour have also been proposed, by extending dipole showers beyond leading-$\Nc$ colour flows \cite{Platzer:2012np,Platzer:2018pmd,Isaacson:2018zdi,Hoeche:2020nsx}. In a recent paper \cite{Forshaw:2020wrq}, we constructed a dipole shower that has the virtue of inheriting some of the colour dynamics of an angular-ordered shower, which improved sub-leading colour accuracy. In this paper we perform a fixed-order cross-check of that approach. We do so by comparing the improved shower's assignment of colour factors to the corresponding exact $e^{+}e^{-}$ matrix elements, computed with second-order QCD corrections. Motivated by these calculations, we are able to further improve our dipole shower's description of colour, in a way that is applicable to evolution with an arbitrary number of emissions.

In \cite{Forshaw:2020wrq} we derived an improved dipole shower in the context of $e^{+}e^{-}\rightarrow q \bar{q}$ collisions\footnote{Though the framework to extend the shower beyond $e^{+}e^{-}$ was presented in the appendices of \cite{Forshaw:2020wrq}.}, starting from an algorithm for the evolution of QCD amplitudes first presented in \cite{Forshaw:2019ver}. The shower can be understood by considering a few key features of angular-ordered and dipole showers. When a shower emits a parton, three new degrees of freedom (DoF) are introduced, describing the new parton's energy and direction. Angular-ordered showers average over one of the DoF (a contextually defined azimuth) which allows the effects of QCD coherence to become manifest. In turn, this reduces the shower to a Markovian sequence of parton decays ($1 \rightarrow 2$ transitions). Thus the final-state partons produced by the shower have a unique branching history with colour factors assigned in accordance with QCD coherence. The angular-ordered approach is very powerful; by harnessing QCD coherence, NLL resummation can be achieved for a broad class of global observables \cite{Resum_large_logs_ee}. However, averaging a DoF limits the approach.

In contrast to angular ordering, the dipole approach retains full dependence on the DoF of each emitted parton. Instead it approximates the colour structures in the shower by emitting partons from colour-connected dipoles. This restricts a basic dipole shower to leading-colour accuracy. Thus a dipole shower is built from a Markovian sequence of $2 \rightarrow 3$ transitions and, as a result, dipole showers lack a unique branching history of parent partons and their decay products. However, a branching history can be constructed by introducing a dipole partitioning, which probabilistically assigns the emitted parton to one of the two parent partons in the dipole. Modern dipole showers use this partitioning to assign colour factors and momentum conservation, and to facilitate hard-process matching. In effect, our approach in \cite{Forshaw:2020wrq} was to define a partitioning so that, after averaging over azimuths, each branching history and its relative weight matches with a corresponding branching history generated by an angular-ordered shower. Through this link, we could assign colour factors beyond leading colour in the dipole shower. As we show in this paper, when applied naively (as was done in \cite{Forshaw:2020wrq}) this procedure does not completely eliminate sub-leading colour errors in the dipole shower for some observables, even at LL accuracy. The problem arises since the $k_t$-ordered dipole shower necessarily involves branching histories disordered in angle: soft, large-angle emissions can appear anywhere in the branching history. These particular branching histories complicate any attempt to assign colour factors in a dipole shower (a point previously noted in \cite{Dasgupta:2018nvj}) and were not completely accounted for in our original approach. In this paper, we solve this problem by introducing dynamical colour factors, i.e. we fix the LL, sub-leading colour errors in event shape observables and increase the shower's sensitivity to full-colour NLLs (falling short of full-colour NLL resummation).

The rest of the paper is structured as follows. After a review of the double emission matrix element in Section \ref{sec:recap}, we repeat the calculation for our original dipole shower in Section \ref{sec:3} and compare the two. We find that the shower works as intended, i.e. the colour factors assigned to partons whose emissions are ordered in angle agree with those of the fixed-order result. However, for emissions unordered in angle, the shower has only leading-colour accuracy.  The understanding brought about by the fixed-order analysis allows us to construct a new method for the correct assignment of dynamical colour factors for emissions unordered in angle. The specific partitioning we introduced in \cite{Forshaw:2020wrq} plays a crucial role in the construction of the new colour factors. The approach we take involves altering shower kernels by introducing a dynamic colour factor that is a function of the branching history. This method involves a computational complexity that asymptotically grows logarithmically with the parton multiplicity. Finally, to illustrate the importance of the dipole partitioning, we compute the $\mathcal{O}(\As^{2})$ difference between exact squared matrix elements and those calculated using a dipole shower employing a different (Catani-Seymour \cite{Catani:1996vz}) partitioning. We find that, in the limit the emissions are strongly ordered in energy and angle, the $\mathcal{O}(\As^{2})$ difference does not vanish, with the possibility of a LL, sub-leading colour error, as was noted in \cite{Dasgupta:2018nvj}. For specific observables (e.g. thrust) this error may be removed at order $\mathcal{O}(\As^{2})$ from a dipole shower employing a Catani-Seymour-type dipole partitioning by using our dynamic colour factors. However, the error will likely re-emerge at higher-orders.

\section{A recap of the $\mathcal{O}(\As^{2})$ QCD squared matrix element}
\label{sec:recap}

\begin{figure}[h]
	\centering
	\includegraphics[width=0.3\textwidth]{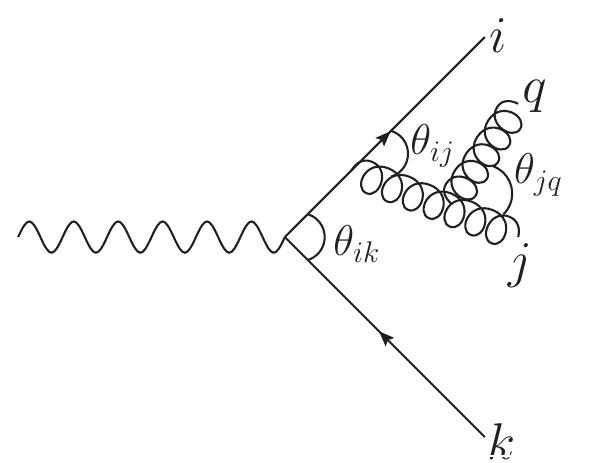}
	\caption{One of the Feynman diagrams contributing to the $\mathcal{O}(\As^{2})$ $e^{+}e^{-}\rightarrow q \bar{q} gg$ matrix element used to compute Eq.~\eqref{eq:matrixelement}. In the present work we calculate these in the soft approximation for which the second gluon is assumed to have energy much less than the first emission.}	\label{fig:matrixlement}
\end{figure}

First we recap the calculation of the $\mathcal{O}(\As^{2})$ $e^{+}e^{-}\rightarrow q \bar{q} gg$ squared matrix element when the gluons are either soft or collinear. Figure~\ref{fig:matrixlement} illustrates our labelling of the partons and the angles between them. This calculation is essentially a recap of Section 5.5 in Ellis, Stirling and Webber \cite{ellis_stirling_webber_1996} and Chapter 4 in Dokshitzer, Khoze, Mueller and Troyan \cite{Dokshitzer:1991wu}. To start, we will only take the limit that lab-frame energies satisfy $E_{q} \ll  E_{j}, E_{i}, E_{k}$ (i.e. the pure soft limit for $q$). Thus our starting point is\footnote{In this case, as the three-parton matrix element is diagonal in colour, we have $\langle {\cal M}_1(...)|{\mathbf T}_i\cdot {\mathbf T}_j|{\cal M}_1(...)\rangle = {\rm Tr}\left[{\mathbf T}_i\cdot {\mathbf T}_j\right] |\mathcal{M}_{1}(...)|^{2}$.}
\begin{align}
|\mathcal{M}_{2}|^{2} \frac{\td^{3} \vec{p}_{q}}{2E_{q}} \approx & -\frac{2\As}{\pi}\Tr \left(\cT_{i}\cdot \cT_{j}w_{ij} + \cT_{j}\cdot \cT_{k}w_{jk} \right. \\ \nonumber
&\left. + \cT_{k}\cdot \cT_{i}w_{ki}\right) \frac{\td E_{q}}{E_{q}} \frac{\td \Omega_{q}}{4\pi} |\mathcal{M}_{1}(\vec{p}_{i},\vec{p}_{j},\vec{p}_{k})|^{2} , \label{eq:matrixelement}
\end{align}
where
\begin{align}
w_{ab} =  \frac{E_{q}^{2}\; p_{a}\cdot p_{b}}{p_{a}\cdot p_{q} \, p_{b} \cdot p_{q}}.
\end{align}
and where $|\mathcal{M}_{1}(\vec{p}_{i},\vec{p}_{j},\vec{p}_{k})|^{2}$ is the $\mathcal{O}(\As)$ squared matrix element. At leading colour (LC) we have
\begin{align}
&|\mathcal{M}_{2}|^{2} \frac{\td^{3} \vec{p}_{q}}{2E_{q}} \nonumber \\
&\quad \approx \frac{\As\Nc}{\pi} \left(  w_{ij} + w_{jk} \right) \frac{\td E_{q}}{E_{q}} \frac{\td \Omega_{q}}{4\pi} |\mathcal{M}_{1}(\vec{p}_{i},\vec{p}_{j},\vec{p}_{k})|^{2}. \label{eq:LCmatrixelement}
\end{align}
This can be interpreted as a sum of emissions from two independent dipoles, $(ij)$ and $(jk)$, and is the basic result on which dipole showers and the Banfi-Marchesini-Smye (BMS) equation \cite{Banfi:2004yd} are built, see also the discussion in \cite{SoftEvolutionAlgorithm} for a more detailed analysis in the case of more general processes.

\begin{figure}[h]
	\centering
	\subfigure[$\Tr \, \cT_{i}^{2} \; P_{ij}^{[i]}$]{\includegraphics[width=0.2\textwidth]{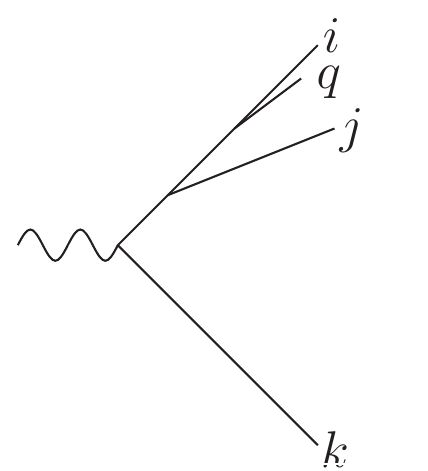}}
	\subfigure[$\Tr  \, \cT_{j}^{2} \; P_{ji}^{[j]}$]{\includegraphics[width=0.2\textwidth]{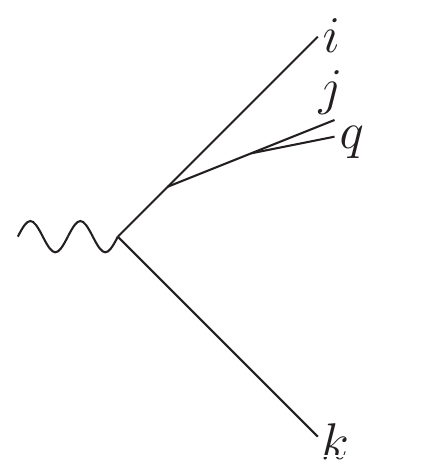}} 	
	\subfigure[$\Tr  \, \cT_{k}^{2} \; P_{k(ij)}^{[k]} $]{\includegraphics[width=0.2\textwidth]{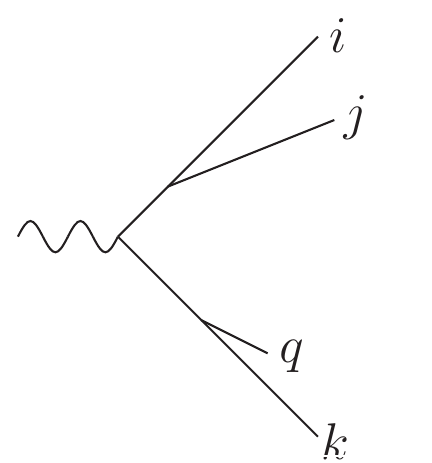}}
	\subfigure[$\Tr  \, \cT_{(ij)}^{2} \; \tilde{P}_{(ij)k}^{[(ij)]} $]{\includegraphics[width=0.2\textwidth]{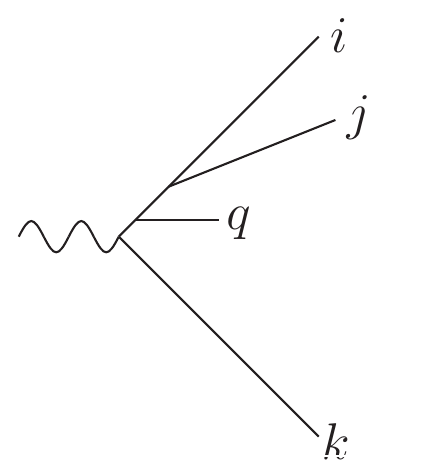}}
	\caption{Diagrams illustrating the angular-ordered interpretation of the four terms in Eq.~\eqref{eq:m2full}. The relative lengths of lines depict the relative energies. Likewise, the relative angles between lines are indicative.}
	\label{fig:intuitionforfixedorder}
\end{figure}

Without approximating colour, we can simplify the matrix element by only keeping terms which are logarithmically enhanced in the two-jet limit (i.e. terms that diverge as $\theta_{ij}/ \theta_{ik} \rightarrow 0$). To do this, we write each $w_{ab}$ as
\begin{align}
w_{ab} =  P_{ab} + P_{ba},
\end{align}
where
\begin{align}
2P_{ab} = w_{ab} + \frac{E_{a}E_{q}}{p_{a}\cdot p_{q}} - \frac{E_{b}E_{q}}{p_{b}\cdot p_{q}}
\end{align}
and
\begin{align}
\int^{2\pi}_{0} \frac{\td \phi^{(a)}_{q}}{2\pi} P_{ab} = \frac{1}{1 - \cos \theta_{aq}} \Theta(\theta_{aq} < \theta_{ab}).
\end{align}
Here $\phi_{q}^{(a)}$ is the azimuth as measured around the direction of $p_{a}$. We define the following shorthand for averaging the azimuths:
\begin{align}
P_{ab}^{[a]} = \frac{(\td\cos\theta_{aq}) \; \td \phi^{(a)}_{q}}{4\pi} \; \int^{2\pi}_{0} \frac{\td \phi^{(a)}_{q}}{2\pi} P_{ab}.
\end{align}
Importantly, $P_{ab}^{[a]}$ only depends on parton $b$ via the theta function constraining the angle of emission. We now consider the limit $\theta_{ij} \ll \theta_{ik} = \pi$, whence we can assume
\begin{align}
&P_{ki}^{[k]} \approx P_{kj}^{[k]} \approx P_{k(ij)}^{[k]},
\end{align}
where $(ij)$ refers to the momentum $p_{(ij)} = p_{i}+p_{j}$, which is approximately on-shell in the collinear limit we consider. By employing this and similar relations we can simplify the matrix element:
\begin{align}
&|\mathcal{M}_{2}|^{2} \frac{\td^{3} \vec{p}_{q}}{2E_{q}} \nonumber \\
& \quad \approx \frac{2\As}{\pi}\Tr \left(\cT_{i}^{2} P_{ij}^{[i]} + \cT_{j}^{2}P_{ji}^{[j]} + \cT_{k}^{2}P_{k(ij)}^{[k]} + \cT_{(ij)}^{2}\tilde{P}_{(ij)k}^{[(ij)]}\right) \nonumber \\
& \qquad \times\frac{\td E_{q}}{E_{q} } |\mathcal{M}_{1}(\vec{p}_{i},\vec{p}_{j},\vec{p}_{k})|^{2}, \label{eq:m2full}
\end{align}
where $\tilde{P}_{(ij)k}^{[(ij)]} = P_{(ij)k}^{[(ij)]} \Theta(\theta_{(ij)q}>\theta_{ij})$. 
The four contributions are illustrated in Figure \ref{fig:intuitionforfixedorder} and 
Table \ref{tbl:limits} tabulates each term in Eq.~\eqref{eq:m2full} across the entire emission phase-space for $q$. 

\begin{table*}[h!]
\begin{center}
    \begin{tabular}{cc ||c|c|c|c}
        $\frac{\theta_{iq}}{\theta_{ij}}$ & $\frac{\theta_{jq}}{\theta_{ij}}$ & $\Tr \, \cT_{i}^{2} \; P_{ij}^{[i]}$ & $\Tr  \, \cT_{j}^{2} \; P_{ji}^{[j]}$ & $\Tr  \, \cT_{k}^{2} \; P_{k(ij)}^{[k]} $ & $\Tr  \, \cT_{(ij)}^{2} \; \tilde{P}_{(ij)k}^{[(ij)]} $ \\ \hline
        $<1$ & $< 1$ & $C_{\TT{F}} \; P_{ij}^{[i]}$ & $C_{\TT{A}} \; P_{ji}^{[j]}$  & \underline{$ C_{\TT{F}} \; P_{k(ij)}^{[k]} $} & $0 $  \\
        $< 1$ & $> 1$ & \underline{$C_{\TT{F}} \; P_{ij}^{[i]}$}  & \underline{$0$}  & \underline{$C_{\TT{F}} \; P_{k(ij)}^{[k]} $} & \underline{$0$}  \\
        $> 1$ & $< 1$ & \underline{$0$} & \underline{$C_{\TT{A}} \; P_{ji}^{[j]}$}  & \underline {$C_{\TT{F}} \; P_{k(ij)}^{[k]} $} & \underline{$0$}  \\
        $> 1$ & $> 1$ & \underline{$0$}  & \underline{$0$} & $C_{\TT{F}} \; P_{k(ij)}^{[k]}$ & \underline{$C_{\TT{F}} \; \tilde{P}_{(ij)k}^{[(ij)]} $} \\
    \end{tabular}
\end{center}
\caption{The contributions to Eq.~\eqref{eq:m2full}, the azimuthally-averaged squared matrix element in the limit $\theta_{ij} \ll \theta_{ik}$. Terms where wide-angle, soft physics has been lost as a result of the collinear approximation are underlined.} \label{tbl:limits}
\end{table*}

It is important to note that when deriving Eq.~\eqref{eq:m2full}, terms such as
\begin{align}
\Delta = \frac{1}{2} \left( P_{ik}^{[i]} - P_{ij}^{[i]} - P_{jk}^{[j]} + P_{ji}^{[j]} \right),
\end{align}
were set to zero by approximating the direction of $i$ and $j$ with a combined direction $(ij)$. For finite $\theta_{ij}$, these terms are only subject to energy divergences and therefore are negligible so long as we insist that a collinear logarithm is picked up. However, because we azimuthally averaged and neglected these pieces, some wide-angle physics is lost which is otherwise captured at LC in Eq.~\eqref{eq:LCmatrixelement} (and consequently in dipole showers and BMS evolution). These soft poles are crucial to a complete description of non-global logarithms. Regions of phase-space for which some wide-angle physics has been set to zero are underlined in Table \ref{tbl:limits}. Crucially, all wide-angle soft physics is included in the limit $\theta_{ij}\rightarrow 0$.

Eq.~\eqref{eq:m2full} can be generalised to include the situation where the parton energies are no longer strongly ordered by introducing hard-collinear physics:
\begin{align}
    \frac{\td E_{q}}{E_{q}} \mapsto \frac{\td E_{q}}{E_{q}} (1+\text{hard-collinear}).
\end{align}
For instance, in the limits defining the rows 1 through 3 of Table \ref{tbl:limits}, Eq.~\eqref{eq:m2full} with hard-collinear physics is
\begin{align}
&|\mathcal{M}_{2}|^{2} \frac{\td^{3} \vec{p}_{q}}{2E_{q}}  \nonumber \\
&\quad \approx \frac{2\As}{\pi} \left(P_{ij}^{[i]}\td z_{i} \; \mathcal{P}_{qq} + P_{ji}^{[j]}\td z_{j} \; \mathcal{P}_{gg} + P_{k(ij)}^{[k]} \td z_{k} \; \mathcal{P}_{qq} \right) \nonumber \\ & \qquad \times |\mathcal{M}_{1}(\vec{p}_{i},\vec{p}_{j},\vec{p}_{k})|^{2}, \label{eq:angleordered}
\end{align}
where $\mathcal{P}_{ab}$ is an unregularised splitting function and $1-z_{m}\approx E_{q}/E_{m}$ with $E_{m}$ the energy of parton $m$ before the emission\footnote{We have ignored the recoil. See Section \ref{sec:momentumcons} for a discussion on its inclusion.}. Likewise, Eq.~\eqref{eq:m2full} in the limit defined in the last row of Table \ref{tbl:limits} becomes
\begin{align}
&|\mathcal{M}_{2}|^{2} \frac{\td^{3} \vec{p}_{q}}{2E_{q}} \nonumber \\ & \quad \approx \frac{2\As}{\pi} \left(P_{(ij)k}^{[(ij)]}\td z_{(ij)} \; \mathcal{P}_{qq} + P_{k(ij)}^{[k]} \td z_{k} \; \mathcal{P}_{qq} \right) \nonumber \\ & \qquad \times |\mathcal{M}_{1}(\vec{p}_{i},\vec{p}_{j},\vec{p}_{k})|^{2}. \label{eq:angleordered2}
\end{align}

\section{Computing the squared matrix element with the dipole shower}

\label{sec:3}

\begin{figure*}[t]
	\centering
	\includegraphics[width=0.88\textwidth]{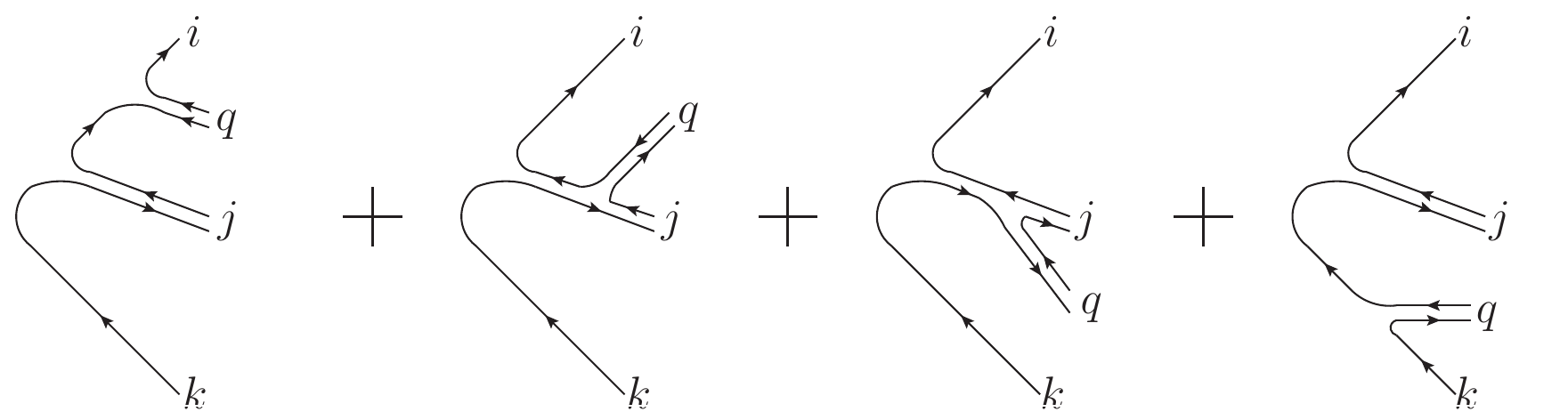}
	\caption{The relevant colour topologies in a dipole shower corresponding to Eq.~\eqref{eq:m2full}. }\label{fig:secondordercolourtopology}
\end{figure*}

Now we want to compute the same squared matrix element using our $k_{t}$-ordered dipole shower. We wish to test that it correctly reproduces the terms in Table \ref{tbl:limits} and the hard-collinear physics in Eqs.~\eqref{eq:angleordered} and \eqref{eq:angleordered2}. The relevant contributions are pictured in Figure~\ref{fig:secondordercolourtopology}.

Consider a generic $k_{t}$-ordered dipole shower, for which emission from a dipole $(a,b)$ at a given slice in $k_{t}$ is generated by an emission probability of
\begin{align}
\frac{\td \TT{Prob}}{\ln k^{ab}_{\bot} } = \int \frac{\td\phi}{2\pi} \frac{\As}{\pi} &\left(C_{a} g_{ab}\mathcal{P}^{\TT{d}}_{a \rightarrow a q}(z^{ab}) \td z^{ab} \right. \nonumber \\
&  \quad \left. + C_{b}g_{ba}\mathcal{P}^{\TT{d}}_{b \rightarrow b q}(z^{ba}) \td z^{ba}\right), \label{eq:Prob}
\end{align}
where $\mathcal{P}^{\TT{d}}_{q \rightarrow q g}(z) = \mathcal{P}^{\TT{d}}_{qq}(z)$ and $\mathcal{P}^{\TT{d}}_{g \rightarrow g g}(z)=\mathcal{P}^{\TT{d}}_{gg}(z)$. We neglect $g\rightarrow q \bar{q}$ transitions, which are sub-leading in colour and only contribute a NLL for doubly-logarithmic observables\footnote{For single-logarithm, collinear-sensitive observables they contribute a leading logarithm at sub-leading colour.}. $\mathcal{P}^{\TT{d}}_{qq}(z)$ and $\mathcal{P}^{\TT{d}}_{gg}(z)$ are the usual dipole  splitting functions, stripped of their colour factors:
\begin{align}
&C_{\TT{F}}\mathcal{P}^{\TT{d}}_{qq}(z) = \mathcal{P}_{qq}(z),\nonumber \\
&\frac{C_{\TT{A}}}{2}\mathcal{P}^{\TT{d}}_{gg}(z) + \frac{C_{\TT{A}}}{2}\mathcal{P}^{\TT{d}}_{gg}(1-z) = \mathcal{P}_{gg}(z),
\end{align}
and where $C_{a} = C_{\TT{F}}$ or $C_{\TT{A}}/2$ if parton $a$ is a quark or gluon. The $g_{ab}$ are dipole partitioning functions, they define how colour factors and momentum conservation  should be distributed across the two members of a dipole and are functions of the momenta of all partons emitted so far. Functions $g_{ab}$ can be smooth or discontinuous functions of the parton momenta. Since we are neglecting momentum conservation in this section, $g_{ab} + g_{ba} = 1$. The relevant kinematic variables are
\begin{align}
&(k^{ab}_{\bot})^{2} = \frac{2 \, p_{a}\cdot p_{q} \, p_{b} \cdot p_{q}}{p_{a} \cdot p_{b}}, \qquad 1-z^{ab} = \frac{p_{q} \cdot p_{b}}{p_{a} \cdot p_{b}}, \nonumber \\ & (k_{\bot})^{2} = \frac{2 \, p_{i}\cdot p_{j} \, p_{k} \cdot p_{j}}{p_{i} \cdot p_{k}},
\end{align}
and $\phi$ is an azimuth so that $\vec{k}^{ab}_{\bot} = k^{ab}_{\bot}(\sin\phi \,  \vec{n}_{1}+\cos\phi \, \vec{n}_{2})$ where $\vec{n}_{1,2}$ are two mutually orthogonal and normalised transverse vectors in the $(a,b)$ dipole zero-momentum frame.

After two emissions, the shower gives
\begin{align}
|\mathcal{M}_{2}|^{2} &\frac{\td^{3} \vec{p}_{q}}{2E_{q}} \approx \nonumber \\ &
\frac{\As}{\pi} \bigg[ \left(C_{\TT{F}} g_{ij}\mathcal{P}^{\TT{d}}_{qq}(z^{ij}) \td z^{ij} + \frac{C_{\TT{A}}}{2}g_{ji}\mathcal{P}^{\TT{d}}_{gg}(z^{ji}) \td z^{ji}\right) \nonumber \\  & \quad \times \frac{\td k^{ij}_{\bot}}{k^{ij}_{\bot}} \frac{\td\phi}{2\pi} \Theta(k^{ij}_{\bot} < k_{\bot}) \nonumber \\ 
&+ \left(\frac{C_{\TT{A}}}{2}g_{jk}\mathcal{P}^{\TT{d}}_{gg}(z^{jk}) \td z^{jk} + C_{\TT{F}}g_{kj} \mathcal{P}^{\TT{d}}_{qq}(z^{kj}) \td z^{kj}\right) \nonumber \\ & \quad  \times  \frac{\td k^{jk}_{\bot}}{k^{jk}_{\bot}} \frac{\td \phi}{2\pi}\Theta(k^{jk}_{\bot} < k_{\bot}) \bigg] |\mathcal{M}_{1}(\vec{p}_{i},\vec{p}_{j},\vec{p}_{k})|^{2}. \label{eq:start}
\end{align}

\subsection{$\mathcal{O}(\As^{2})$ with emissions ordered in angle}
\label{sec:angular-orderedemissions}

We will first consider whether our dipole shower can recreate the physics in rows 1 through 3 of Table \ref{tbl:limits}. The diagrams contributing to this limit will be produced in our shower when the parton transverse momenta and angles are concurrently ordered. For now we will neglect recoil and hard-collinear pieces. Keeping only the soft parts, we have
\begin{align}
& |\mathcal{M}_{2}|^{2} \frac{\td^{3} \vec{p}_{q}}{2E_{q}} \approx \nonumber \\ &  \frac{2\As}{\pi} \bigg[ \left(C_{\TT{F}} g_{ij}w_{ij} + \frac{C_{\TT{A}}}{2}g_{ji}w_{ij} \right) \frac{\td E_{q}}{E_{q}} \frac{\td^{2} \Omega_{q}}{4\pi} \Theta(k^{ij}_{\bot} < k_{\bot}) \nonumber \\ 
& + \left(\frac{C_{\TT{A}}}{2}g_{jk} w_{jk} + C_{\TT{F}}g_{kj} w_{jk}\right) \frac{\td E_{q}}{E_{q}} \frac{\td^{2} \Omega_{q}}{4\pi} \Theta(k^{jk}_{\bot} < k_{\bot}) \bigg]\nonumber \\ 
& \qquad \times |\mathcal{M}_{1}(\vec{p}_{i},\vec{p}_{j},\vec{p}_{k})|^{2}.
\end{align}
Our dipole shower is built using the partitioning
\begin{align}
g_{ab} = \frac{1}{2} + \TT{Asym}_{a,b},
\label{eq:partitioning1}
\end{align}
where \cite{Forshaw:2020wrq}
\begin{align}
    &\TT{Asym}_{a,b} = \left[ \frac{T \cdot p_{a}}{4T \cdot p_{q} }\frac{(k^{(ab)}_{\bot})^{2}}{p_{a} \cdot p_{q}} - \frac{T \cdot p_{b}}{4T \cdot p_{q} }\frac{(k^{(ab)}_{\bot})^{2}}{p_{b} \cdot p_{q}}\right], \nonumber \\ & \TT{and} \quad T = \sum_{i} p_{i},
\end{align}
where the sum over $i$ is a sum over all partons in the event. $T$ plays the role of projecting the lab frame energy when it is contracted with a momentum vector. Roughly speaking, this way of partitioning a dipole corresponds to splitting the dipole in half in the laboratory frame; it is defined specifically to ensure
\begin{align}
g_{ab} \, w_{ab} = P_{ab}, \label{eq:partitioning2}
\end{align}
and therefore
\begin{align}
|\mathcal{M}_{2}|^{2}& \frac{\td^{3} \vec{p}_{q}}{2E_{q}}  \approx \nonumber \\ & \frac{2\As}{\pi} \bigg[ \left(C_{\TT{F}} P_{ij} + \frac{C_{\TT{A}}}{2}P_{ji} \right)\frac{\td E_{q}}{E_{q}} \frac{\td^{2} \Omega_{q}}{4\pi} \Theta(k^{ij}_{\bot} < k_{\bot}) \nonumber \\ 
&+ \left(\frac{C_{\TT{A}}}{2} P_{jk} + C_{\TT{F}} P_{kj}\right) \frac{\td E_{q}}{E_{q}} \frac{\td^{2} \Omega_{q}}{4\pi} \Theta(k^{jk}_{\bot} < k_{\bot}) \bigg] \nonumber \\ & \qquad \times |\mathcal{M}_{1}(\vec{p}_{i},\vec{p}_{j},\vec{p}_{k})|^{2}.\label{eq:othermidstep}
\end{align}
After azimuthal averaging:
\begin{align}
|\mathcal{M}_{2}|^{2} &\frac{\td^{3} \vec{p}_{q}}{2E_{q}} \approx \nonumber \\ 
&\frac{2\As}{\pi} \bigg[\left(C_{\TT{F}} P^{[i]}_{ij} + \frac{C_{\TT{A}}}{2}P^{[j]}_{ji} \right)\frac{\td E_{q}}{E_{q}} \Theta(k^{ij}_{\bot} < k_{\bot}) \nonumber \\ 
&+ \left(\frac{C_{\TT{A}}}{2} P^{[j]}_{jk} + C_{\TT{F}} P^{[k]}_{kj}\right) \frac{\td E_{q}}{E_{q}} \Theta(k^{jk}_{\bot} < k_{\bot}) \bigg] \nonumber \\ 
& \qquad \times |\mathcal{M}_{1}(\vec{p}_{i},\vec{p}_{j},\vec{p}_{k})|^{2}. \label{eq:midstep}
\end{align}
As we are working in the limits defined in rows 1 through 3 of Table \ref{tbl:limits}, and in the soft limit for $q$, the $k_{t}$ ordering theta functions all saturate and can be removed. Thus, we find 
\begin{align}
|\mathcal{M}_{2}|^{2} \frac{\td^{3} \vec{p}_{q}}{2E_{q}} \approx &\frac{2\As}{\pi} \left(C_{\TT{F}} P_{ij}^{[i]} + C_{\TT{A}}P_{ji}^{[j]} + C_{\TT{F}}P_{k(ij)}^{[k]} \right) \nonumber \\ & \times \frac{\td E_{q}}{E_{q} } |\mathcal{M}_{1}(\vec{p}_{i},\vec{p}_{j},\vec{p}_{k})|^{2}. \label{eq:dipolewithouthard}
\end{align}
This is the same as the fixed-order result given in the previous section. 

We will now relax the soft approximation and check whether our dipole shower correctly includes the hard-collinear physics too, i.e. that it reconstructs Eq.~\eqref{eq:angleordered}. We can once again start from Eq.~\eqref{eq:start}: 
\begin{align}
&C_{a} g_{ab}\mathcal{P}^{\TT{d}}_{a \rightarrow a q}(z^{ab}) \frac{\td k^{ab}_{\bot}}{k^{ab}_{\bot}} \td z^{ab} \frac{\td\phi}{2\pi}\nonumber \\ &  = C_{a} g_{ab} w_{ab} \frac{\td E_{q}}{E_{q}} \frac{\td^{2} \Omega_{q}}{2\pi} (1+\TT{hard \; pieces}) \nonumber \\ &= C_{a} P_{ab} \frac{\td E_{q}}{E_{q}} \frac{\td^{2} \Omega_{q}}{2\pi} (1+\TT{hard \; pieces}),
\end{align}
where the `hard pieces' part depends on the splitting function:
\begin{align}
&\mathcal{P}^{\TT{d}}_{a \rightarrow a q} = \mathcal{P}^{\TT{d}}_{q q} : \quad \TT{hard \; pieces} = (z^{ab})^{2}, \nonumber \\
&\mathcal{P}^{\TT{d}}_{a \rightarrow a q} = \mathcal{P}^{\TT{d}}_{gg} : \quad \TT{hard \; pieces} = (z^{ab})^{3}.
\end{align}
Since the collinear limit requires $\theta_{iq} \text{~or~} \theta_{jq} \ll \theta_{ij} \ll \theta_{ik}$, we can let $z^{ab} \approx z_{a}$ where $1-z_{a}\approx E_{q}/E_{a}$ and where $E_{a}$ is the energy of parton $a$ before the emission is generated. Thus `hard pieces' does not have any azimuthal dependence\footnote{`hard pieces' do contain azimuthal dependence if we include spin correlations. In \cite{Forshaw:2019ver}, we  discussed using the Collins and Knowles algorithm \cite{Collins:1987cp,KNOWLES1990271} to re-introduce spin correlations by re-weighting after the shower has terminated.} and so azimuthal averaging proceeds as before. We find that
\begin{align}
& |\mathcal{M}_{2}|^{2} \frac{\td^{3} \vec{p}_{q}}{2E_{q}} \approx \nonumber \\ & \frac{2\As}{\pi}  \bigg(\left[P_{ij}^{[i]}\td z_{i} \; \mathcal{P}_{qq} + P_{ji}^{[j]}\td z_{j} \; \frac{C_{\TT{A}}}{2} \mathcal{P}^{d}_{gg}(z_{j})\right] \Theta(k^{ij}_{\bot} < k_{\bot})  \nonumber \\ 
& + \left[P_{jk}^{[j]}\td z_{j}\frac{C_{\TT{A}}}{2} \mathcal{P}^{d}_{gg}(1-z_{j}) + P_{k(ij)}^{[k]} \td z_{k} \; \mathcal{P}_{qq}\right] \nonumber \\ & \qquad  \times \Theta(k^{jk}_{\bot} < k_{\bot}) \bigg) |\mathcal{M}_{1}(\vec{p}_{i},\vec{p}_{j},\vec{p}_{k})|^{2}. \label{eq:dipolewithhard}
\end{align}
Once again, the collinear limit results in $P_{ji}^{[j]} = P_{jk}^{[j]}$. The small-angle approximation for $q$ saturates the ordering theta functions and so 
\begin{align}
&|\mathcal{M}_{2}|^{2} \frac{\td^{3} \vec{p}_{q}}{2E_{q}} \approx \nonumber \\ & \frac{2\As}{\pi} \bigg(P_{ij}^{[i]}\td z_{i} \; \mathcal{P}_{qq}  + P_{ji}^{[j]}\td z_{j} \; \mathcal{P}_{gg}(z_{j}) + P_{k(ij)}^{[k]} \td z_{k} \; \mathcal{P}_{qq} \bigg)\nonumber \\ & \qquad \times |\mathcal{M}_{1}(\vec{p}_{i},\vec{p}_{j},\vec{p}_{k})|^{2}. \label{eq:3.13}
\end{align}
This is equivalent to the fixed-order result of Eq.~\eqref{eq:angleordered}.

An important part of this section was to assume we can neglect recoil and that further emissions do not modify momenta in such a way that these correctly computed matrix elements are destroyed. As we showed explicitly in Section 3.1 of \cite{Forshaw:2020wrq}, our global recoil does not mess the computation of NLLs at this order. This is further discussed in Section \ref{sec:momentumcons}.

\subsection{$\mathcal{O}(\As^{2})$ with emissions unordered in angle}

In the previous section we validated our dipole shower's ability to reproduce rows 1 through 3 of Table \ref{tbl:limits}. In this section we wish to test the shower's ability to reproduce the last row of Table \ref{tbl:limits} and the LC limit in Eq.~\eqref{eq:LCmatrixelement}, which is applicable across all the limits considered in the table and also when $\theta_{ij} \not\ll \theta_{ik}$. To start we will test our dipole shower in the limit that $q$ is soft whilst $\theta_{ij} \ll \theta_{ik}$ but $\theta_{iq} \approx \theta_{jq} > \theta_{ij}$ (i.e. we will compare against row 4 in Table \ref{tbl:limits}). We describe these emissions as unordered in angle since they are produced in the shower with angles out of order; the $k_{t}$ and angle of these are emissions are not concurrently ordered. However, these emissions can still have a strong angular hierarchy allowing them to produce a LL, i.e. $\theta_{ij} \ll \theta_{iq} \approx \theta_{jq} \ll  \theta_{ik}$. The region of phase-space which has this hierarchy is highly restricted, due to the opposing $k_{t}$ ordering, but is nevertheless present and its mistreatment can induce a (small) LL error in some observables, for instance thrust \cite{Dasgupta:2018nvj}. At the end of this section we will check the crucial soft, wide-angle limit, where parton $q$ is soft but all angles are unconstrained.

We will begin from Eq.~\eqref{eq:midstep}, which was derived from our shower by only assuming parton $q$ is soft. Employing $\theta_{ij} \ll \theta_{ik}$ allows us to replace $P_{ki}^{[k]} \approx P_{kj}^{[k]} \approx P_{k(ij)}^{[k]}$, i.e.
\begin{align}
&|\mathcal{M}_{2}|^{2} \frac{\td^{3} \vec{p}_{q}}{2E_{q}} \approx \frac{2\As}{\pi} \bigg[ \left(C_{\TT{F}} P^{[i]}_{ij} + \frac{C_{\TT{A}}}{2}P^{[j]}_{ji} \right)\frac{\td E_{q}}{E_{q}} \Theta(k^{ij}_{\bot} < k_{\bot}) \nonumber \\ 
&+ \left(\frac{C_{\TT{A}}}{2} P^{[j]}_{jk} + C_{\TT{F}}P_{k(ij)}^{[k]}\right) \frac{\td E_{q}}{E_{q}} \Theta(k^{jk}_{\bot} < k_{\bot}) \bigg] |\mathcal{M}_{1}(\vec{p}_{i},\vec{p}_{j},\vec{p}_{k})|^{2}. \label{eq:3.14}
\end{align}
Now we take the limit that $\theta_{iq},\theta_{jq} \gg \theta_{ij}$, thus $P^{[i]}_{ij} = P^{[j]}_{ji} = 0$ and $P^{[j]}_{jk} \approx P^{[(ij)]}_{(ij)k}$ which gives rise to
\begin{align}
&|\mathcal{M}_{2}|^{2} \frac{\td^{3} \vec{p}_{q}}{2E_{q}} \approx \frac{2\As}{\pi} \left(\frac{C_{\TT{A}}}{2} P^{[(ij)]}_{(ij)k} + C_{\TT{F}}P_{k(ij)}^{[k]}\right) \nonumber \\ & \qquad \qquad  \times \frac{\td E_{q}}{E_{q}} \Theta(k^{(ij)k}_{\bot} < k_{\bot}) |\mathcal{M}_{1}(\vec{p}_{i},\vec{p}_{j},\vec{p}_{k})|^{2}.
\end{align}
We should add to this the contribution where parton $q$ is emitted first. Doing so gives
\begin{align}
&|\mathcal{M}_{2}|^{2} \frac{\td^{3} \vec{p}_{q}}{2E_{q}} \approx  \nonumber \\
&\frac{2\As}{\pi} \Bigg[\left(\frac{C_{\TT{A}}}{2}\Theta(k^{(ij)k}_{\bot} < k_{\bot}) + C_{\TT{F}} \Theta(k^{(ij)k}_{\bot} > k_{\bot})\right)P^{[(ij)]}_{(ij)k} \nonumber \\ & \qqquad + C_{\TT{F}}P_{k(ij)}^{[k]}\Bigg] \frac{\td E_{q}}{E_{q}} |\mathcal{M}_{1}(\vec{p}_{i},\vec{p}_{j},\vec{p}_{k})|^{2}. \label{eq:unorderedinangle}
\end{align}
Comparing with row 4 of Table \ref{tbl:limits} we see an $\Nc^{-2}$ suppressed error in the colour factor of the $P^{[(ij)]}_{(ij)k}$ term. This error is due to the parton angles being disordered and is not present in an angular-ordered shower. However, in our dipole shower the disordered configuration is present and important, since it is required to get the correct wide-angle, soft physics beyond the two-jet limit. 

Following the same logic as before, it is simple to show that our dipole shower includes the hard-collinear physics in Eq.~\eqref{eq:angleordered2} with the same $\Nc^{-2}$ suppressed error as in Eq.~\eqref{eq:unorderedinangle}. 

Finally, a good dipole shower should encode Eq.~\eqref{eq:LCmatrixelement} and therein BMS evolution in the limit that the emission is soft. Starting from Eq.~\eqref{eq:othermidstep} and taking the leading colour limit:
\begin{align}
& |\mathcal{M}_{2}|^{2} \frac{\td^{3} \vec{p}_{q}}{2E_{q}} \approx  \frac{\As \Nc}{\pi} \bigg[ \left(P_{ij} + P_{ji} \right)\frac{\td E_{q}}{E_{q}} \frac{\td^{2} \Omega_{q}}{4\pi} \Theta(k^{ij}_{\bot} < k_{\bot}) \nonumber \\ 
&+ \left(P_{jk} + P_{kj}\right) \frac{\td E_{q}}{E_{q}} \frac{\td^{2} \Omega_{q}}{4\pi} \Theta(k^{jk}_{\bot} < k_{\bot}) \bigg] |\mathcal{M}_{1}(\vec{p}_{i},\vec{p}_{j},\vec{p}_{k})|^{2}.
\end{align}
This is equal to
\begin{align}
|\mathcal{M}_{2}|^{2} \frac{\td^{3} \vec{p}_{q}}{2E_{q}} \approx & \frac{\As \Nc}{\pi} \bigg[ w_{ij}  \Theta(k^{ij}_{\bot} < k_{\bot}) + w_{jk}  \Theta(k^{jk}_{\bot} < k_{\bot}) \bigg] \nonumber \\ & \times  \frac{\td E_{q}}{E_{q}} \frac{\td^{2} \Omega_{q}}{4\pi} |\mathcal{M}_{1}(\vec{p}_{i},\vec{p}_{j},\vec{p}_{k})|^{2}, \label{eq:dipoleLC}
\end{align}
which is equal to Eq.~\eqref{eq:LCmatrixelement}, modulo the use of $k_{t}$ instead of energy as the ordering variable, which does not hinder the logarithmic accuracy \cite{ColoumbGluonsOrdering,ColoumbGluonsOrderingLetter}. Hence the dipole shower correctly handles wide-angle soft radiation in the LC approximation. To go beyond the LC approximation generally requires amplitude-level evolution \cite{SoftEvolutionAlgorithm,Forshaw:2019ver,DeAngelis:2020rvq,Nagy:2015hwa,Nagy:2017ggp,Nagy:2019pjp}. 

\subsection{Summary}

\begin{figure*}[t]
	\centering
	\subfigure[]{\includegraphics[width=0.5\textwidth]{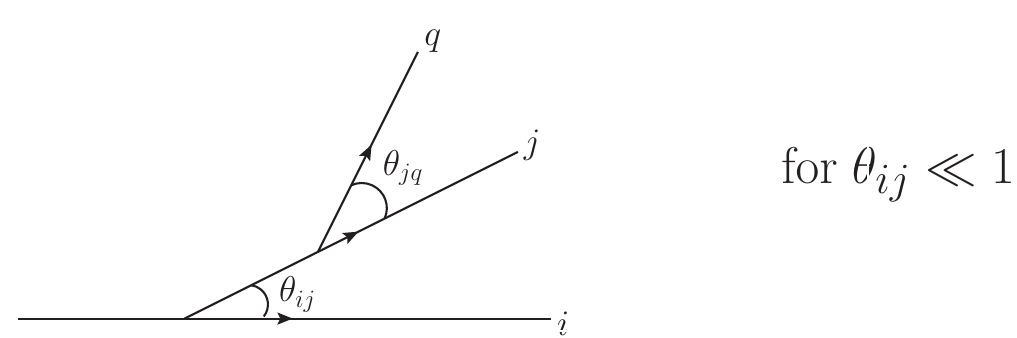}}
	\subfigure[]{\includegraphics[width=\textwidth]{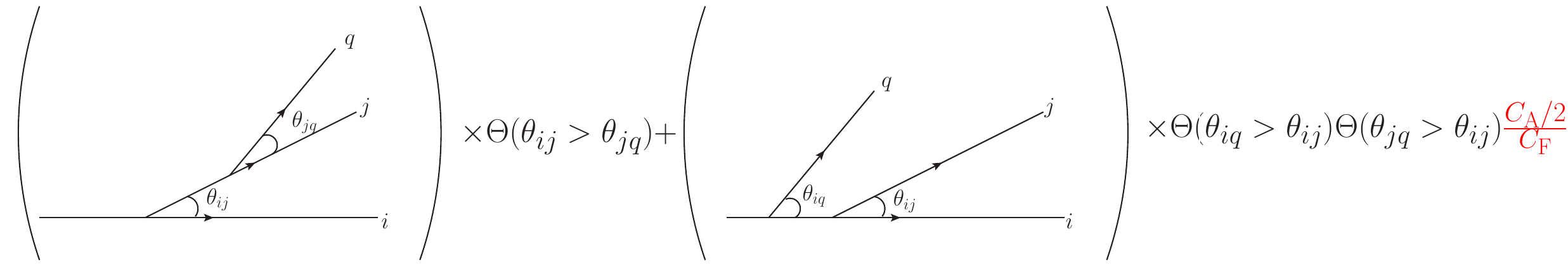}}
	\caption{Diagram (a) is generated using our dipole shower, after partitioning. This topology is where the $\Nc^{-2}$ error emerges. Diagram (b) represents the re-arrangements of (a) that can be made in the limit $\theta_{ij} \ll 1$. These diagrams correspond to those of an angular-ordered shower. The red factor is the $\Nc^{-2}$ suppressed error produced by our original dipole shower.}\label{fig:dipoleerrordiagram}
\end{figure*}

In this section we have evaluated the accuracy at which our original dipole shower recreates the squared matrix elements summarised in Section \ref{sec:recap}. In summary, when parton $q$ is emitted from parton $i$ or $k$ the matrix element is reproduced without error. When parton $q$ is emitted from parton $j$ there is an $\Nc^{-2}$ suppressed error. We can look at different limits of the phase-space for partons $j$ and $q$ and evaluate the colour accuracy of our shower in each limit as follows:
\begin{enumerate}
    \item $\theta_{ij} \ll 1$:
        \begin{enumerate}
            \item $\theta_{jq} \ll \theta_{ij}$: in this region an angular-ordered shower has full colour accuracy and our shower agrees with an angular-ordered shower (see rows 1 and 3 of Table \ref{tbl:limits}).
            \item $\theta_{jq} \sim \theta_{ij}$: in this region an angular-ordered shower cannot recreate the complete matrix element and our shower only guarantees LC accuracy in the soft limit. This region does not contain a strong angular hierarchy so at most can contribute a NLL and is suppressed further in event shape observables only sensitive to perturbations from the two-jet limit, for instance thrust.
            \item $\theta_{jq} \approx \theta_{iq} \gg \theta_{ij}$: in this region an angular-ordered shower has full colour accuracy and our shower currently lacks complete agreement with an angular-ordered shower beyond LC (row 4 of Table \ref{tbl:limits}). This is the region we will address in Section \ref{sec:correcting}.
        \end{enumerate}
    \item $\theta_{ij} \sim 1$: angular ordering cannot describe this region and our shower only guarantees LC accuracy.\footnote{In this limit (which is potentially subject to all manner of soft and non-global logarithms), it is difficult to make statements on the logarithmic accuracy of the shower beyond the leading accuracy of soft logarithms achievable through the BMS equation \cite{BMSEquation}, which is embedded in the dipole shower approach. Though, with this in mind, Dasgupta et al. \cite{Dasgupta:2020fwr} have demonstrated LC NLL accuracy in non-global observables for dipole showers with carefully constructed global recoils and lab-frame based dipole partitionings. Our shower has both these properties and our fixed-order tests of the shower \cite{Forshaw:2020wrq} are consistent with their results. Note that Dasgupta et al.'s definition of NLL accuracy encompasses NLL in the exponent but is also applicable to logs that do not resum into an exponential form such as non-global logs.}
\end{enumerate}
In Figure \ref{fig:dipoleerrordiagram} we illustrate the origin of the $\Nc^{-2}$ suppressed error: the erroneous factor is shown in red. 
The diagram in this figure is sufficient to enable us to read off the correct colour factor, and we make heavy use of this perspective in what follows.

\section{Colour factors for emissions unordered in angle}
\label{sec:correcting}

In the previous section we computed the double-emission  matrix elements squared corresponding to $e^+ e^- \to q \bar{q} gg$, comparing the result from our dipole shower formalism with the relevant limits of the exact matrix element. We showed that, when the two emissions are strongly ordered in angle (with one emission collinear in the direction of one of the hemispheres), the matrix elements calculated from our dipole shower were correct except when a gluon is emitted with an angle larger than the opening angle of its parent dipole. In such a configuration, the coherent branching calculation would correctly assign a colour factor $C_F$, whilst the dipole shower gives $C_{\TT{A}}/2$ (see Eq.~\eqref{eq:unorderedinangle}). At this order, we can correct the colour factor by replacing $C_{i}$ in Eq.~\eqref{eq:Prob} with a dynamic colour factor of
\begin{align}
    &\mathcal{C}_{i j}(\theta_{i q}, \theta_{ij}) = \left(C_{F}\delta^{(q)}_{i} + \frac{C_{A}}{2}\delta^{(g)}_{i} \right)\theta(\theta_{i q} < \theta_{ij}) \nonumber \\
    &+ \left(\frac{C_{A}}{2}(\delta^{(q)}_{i}\delta^{(q)}_{j} + \delta^{(g)}_{i}\delta^{(g)}_{j})+ C_{F}(\delta^{(q)}_{i}\delta^{(g)}_{j} + \delta^{(g)}_{i}\delta^{(q)}_{j}) \right)\nonumber \\ & \qquad \times \theta(\theta_{i q} > \theta_{ij}), \label{eq:firstfix}
\end{align}
where $\delta^{(q)}_{i}$ ($\delta^{(g)}_{i}$) is one when the parton $i$ is a quark (gluon), and zero otherwise.  We stress that this correction leads to the correct result only because our way of partitioning is able to encode angular ordering via
\begin{align}
\frac{(\td\cos\theta_{aq}) \; \td \phi^{(a)}_{q}}{4\pi}  \; \int^{2\pi}_{0} \frac{\td \phi^{(a)}_{q}}{2\pi} g_{ab} \, w_{ab} = P_{ab}^{[a]}, \label{eq:importantcriteria}
\end{align}
which ensures that the error is localized in the colour factor of Eq.~\eqref{eq:unorderedinangle}.  Our partitioning satisfies this requirement exactly.\footnote{In \ref{sec:limitations} we discuss tests for checking whether other partitionings are consistent with the requirement at NLL accuracy.} 

\begin{figure*}[t]
	\centering
	\subfigure[]{\includegraphics[width=0.4\textwidth]{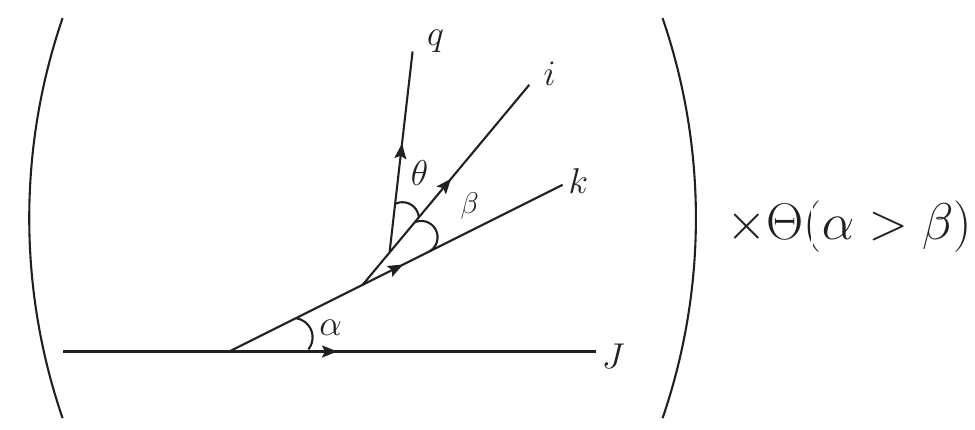}}
	\subfigure[]{\includegraphics[width=\textwidth]{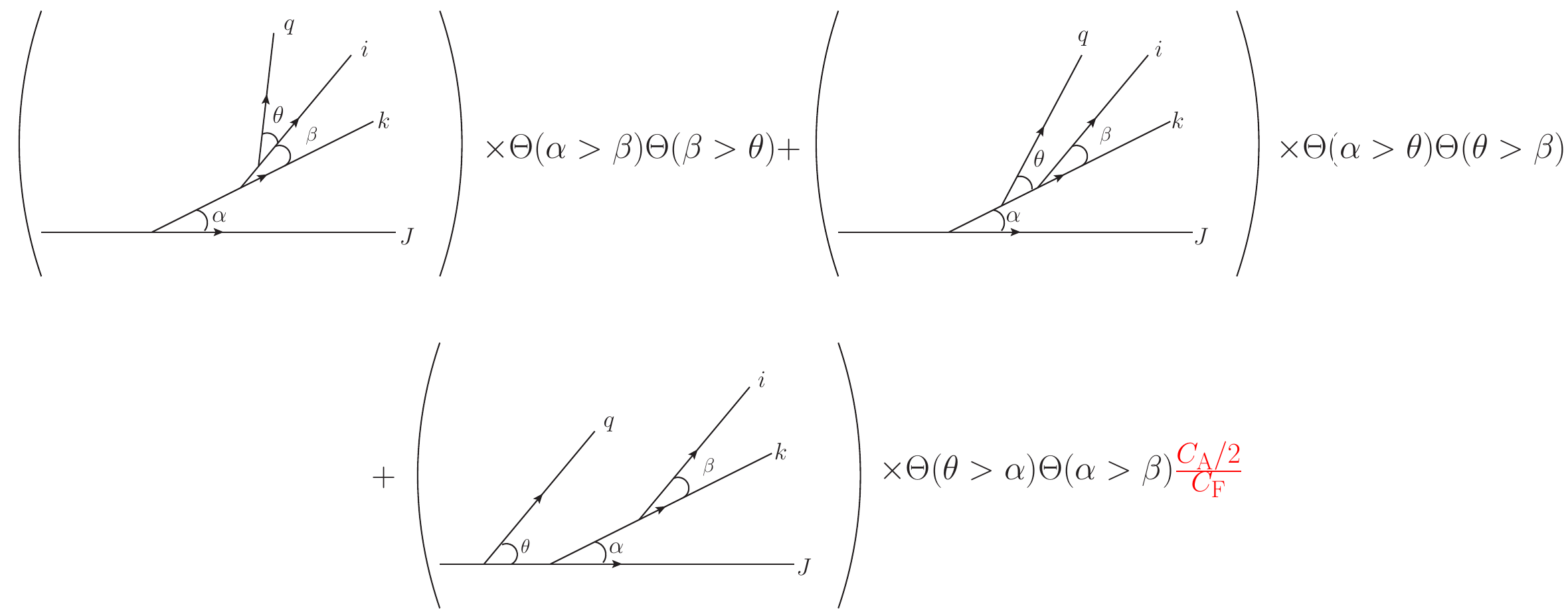}}
	\caption{Diagram (a) is generated by our dipole shower and is chosen because it contains an incorrect colour factor. Diagram (b) represents the re-arrangements of (a) corresponding to an angular-ordered shower. The red colour factor is the $\Nc^{-2}$ suppressed error produced by our original dipole shower.} \label{fig:correctfactors3O1}
\end{figure*}
\begin{figure*}[t]
	\centering
	\subfigure[]{\includegraphics[width=0.4\textwidth]{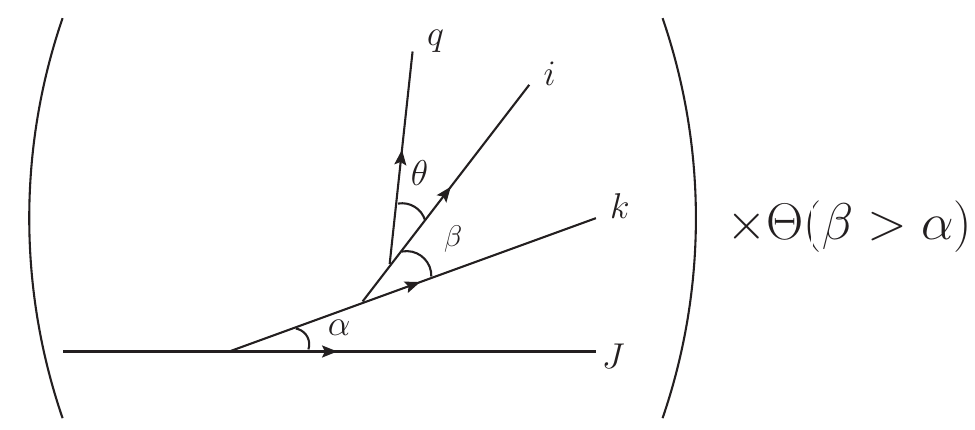}}
	\subfigure[]{\includegraphics[width=\textwidth]{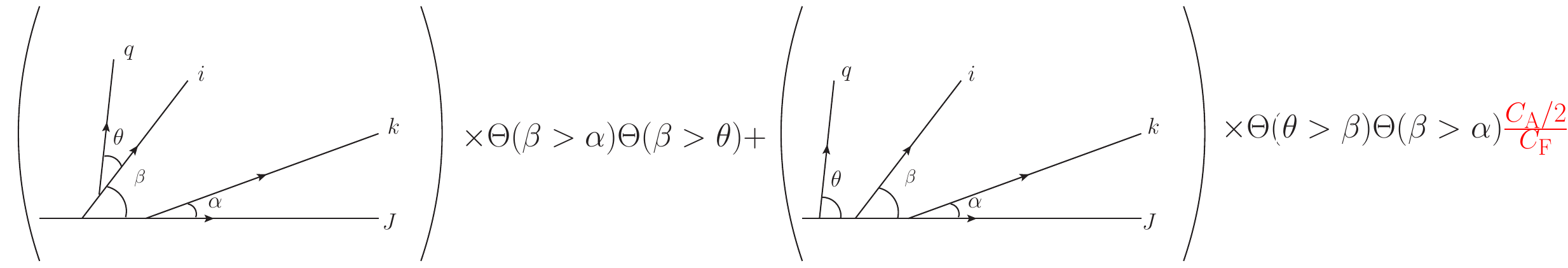}}
	\caption{A second possible ordering of angles that also leads to an incorrect colour factor.} \label{fig:correctfactors3O2}
\end{figure*}

It is not too difficult to generalize to higher orders, and the solution is particularly straightforward in the absence of $g\to q\bar{q}$ branchings, which will be discussed at the end of this section (see also \cite{Friberg:1996xc}). Figures \ref{fig:correctfactors3O1} and \ref{fig:correctfactors3O2} illustrate errors that occur in the case of three emissions. They highlight a key feature: the colour factor of the last emission is incompatible with coherence only when it is emitted at an angle larger than the angular extent of the colour charge distribution of the chain of partons leading to the emission. 

\begin{figure*}[t]
	\centering
	\subfigure[]{\includegraphics[width=0.4\textwidth]{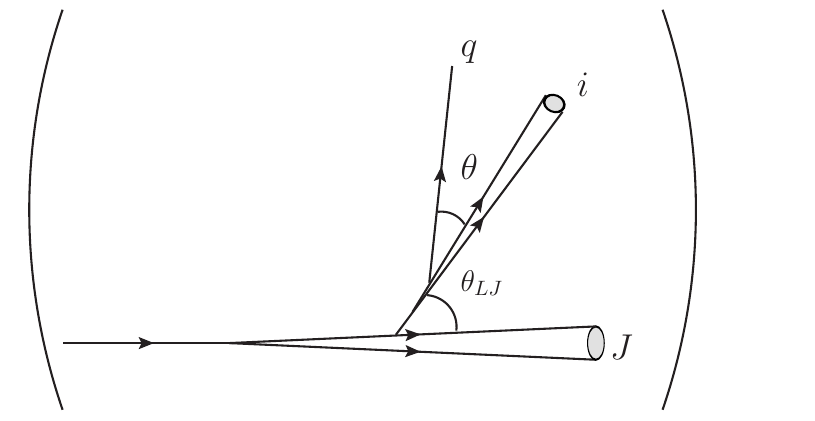}}
	\subfigure[]{\includegraphics[width=\textwidth]{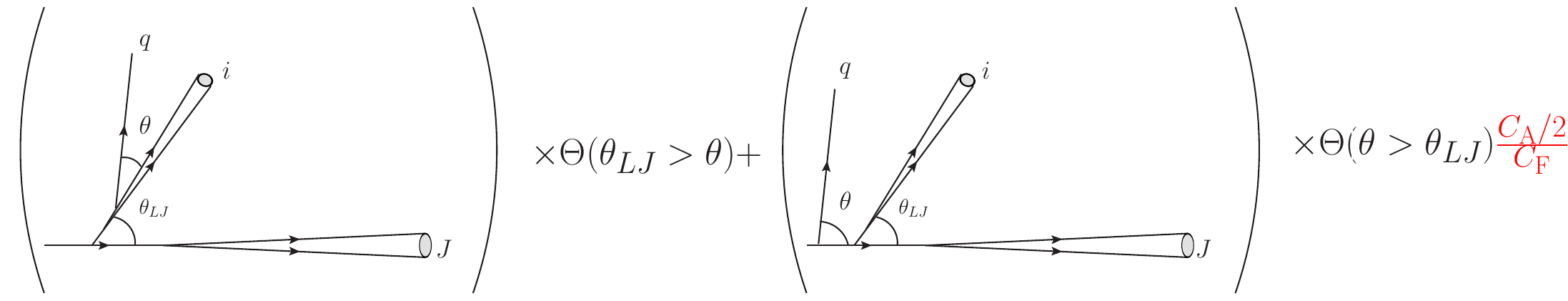}}
	\caption{The generalisation of Figures \ref{fig:correctfactors3O1} and \ref{fig:correctfactors3O2} to an arbitrary fixed order. Cones $i$ and $J$ represent a unspecified number of parton branchings, each at angles smaller than $\theta_{LJ}$, which is the largest angle in $q$'s sub-branch. As before, diagram (a) is generated by our dipole shower and contains an incorrect colour factor associated with the emission of $q$. Diagram (b) represents the re-arrangements of (a) corresponding to an angular-ordered shower. The red colour factor is the $\Nc^{-2}$ suppressed error produced by our original dipole shower.}	\label{fig:correctfactorsAO}
\end{figure*}

\begin{figure}[h]
	\centering
	\includegraphics[width=0.5\textwidth]{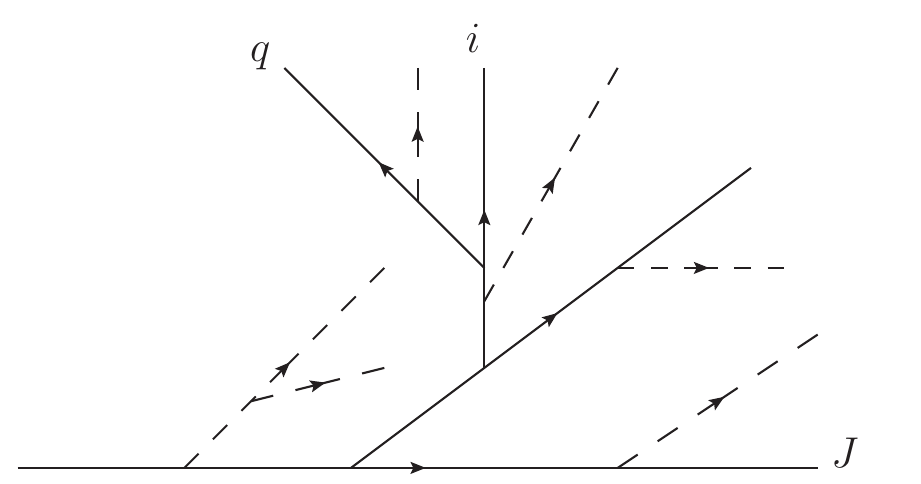}
	\caption{An illustration of a branch containing hard parton $J$. The sub-branch for parton $q$ contains the partons with solid lines, these form parton $q$'s `parental chain'. Partons with dashed lines are in $J$'s branch but are not in $q$'s sub-branch. The sub-branch length is the number of partons in a sub-branch: parton $q$'s sub-branch has a length 4 whilst parton $i$'s sub-branch has a length 3.}	\label{fig:branch}
\end{figure}

Figure \ref{fig:correctfactorsAO} shows the generalisation to an arbitrary fixed order\footnote{In \ref{sec:diagrams} we show that the planar diagrams arising after azimuthal averaging do generalise to higher orders}. As a consequence of using a partitioning which defines a unique branching history of $1\rightarrow 2$ transitions, the collection of partons in an event can be divided into $m$ branches for an $m$ parton hard process. Each branch contains one of the hard-process partons and the radiation emitted from it. Each parton in the branch can also be assigned a unique sub-branch consisting of the parton and its ``parental chain'', see Figure \ref{fig:branch}. We only need to modify colour factors for gluons which cannot probe the largest angle in their sub-branch. We do this by extending the definition of $\mathcal{C}_{i j}$ to
\begin{align}
    \mathcal{C}_{i J}(\theta_{i q},\theta_{L J}) =& \left(C_{F}\delta^{(q)}_{i} + \frac{C_{A}}{2}\delta^{(g)}_{i} \right)\theta(\theta_{i q} < \theta_{L J}) \nonumber \\ & + \left(\frac{C_{A}}{2}\delta^{(g)}_{J}+ C_{F}\delta^{(q)}_{J} \right)\theta(\theta_{i q} > \theta_{L J}), \label{eqn:newcolourfactor}
\end{align}
where $J$ is the hard parton in the sub-branch and $L$ is the parton in the sub-branch emitted at the largest angle. $\theta_{L J}$ is the angle between $L$ and $J$.\footnote{$\theta_{L J}=\pi$ for an emission with a sub-branch of length $2$.} One should use these newly defined dynamic colour factors in both emissions and in the Sudakov form factors, i.e. so that the two are related by unitarity.\footnote{All current dipole shower implementations \cite{Herwig_dipole_shower,Pythia8,Schumann:2007mg} could directly employ our algorithm using existing methods such as the Sudakov veto algorithm.} The computation of the dynamic colour factor grows at most linearly as the shower progresses and on average logarithmically.\footnote{The average sub-branch length for a multiplicity, $n$, of partons in the branch is $\sum^{n}_{i=1}\frac{1}{i} \leq \ln n + 1$.}

In summary we have constructed a dipole shower which encodes the physics of QCD coherence just as in an angular-ordered shower. The resulting dipole shower, at LC, reproduces BMS evolution and, after using the CMW running coupling \cite{Catani:1996vz}, will match the NLL-accurate dipole showers with global recoils discussed in \cite{Dasgupta:2020fwr}. In all, we expect our dipole shower to achieve full colour LL accuracy in any observable for which an angular-ordered shower can also be used to resum LLs. In the case of $e^+ e^- \rightarrow q \bar{q}$, the NLLs of some observables (i.e. thrust) do not directly depend on $g\rightarrow q\bar{q}$ transitions\footnote{Such transitions are restricted to secondary branchings, the remnants of which can be resummed into the CMW coupling and are otherwise rendered trivial by the angular-ordering constraint \cite{Resum_large_logs_ee}.} in which case our shower is accurate to NLL at full colour. Our methodology of assigning colour factors by mapping branching histories onto those of an angular-ordered shower could be generalised to assign the correct colour factors after including $g\rightarrow q\bar{q}$ transitions.\footnote{Furthermore, our arguments also generalise to a hard process with more than two coloured, hard legs provided each of the dipoles found in the colour flow for the hard process is evolved in its back-to-back frame as is done in an angular ordered shower \cite{Herwig_shower}. See Appendix A of \cite{Forshaw:2020wrq} for a more complete discussion on generalising our shower beyond $e^+e^- \rightarrow q \bar{q}$.} However, because these transitions introduce more quark lines into the parton cascade, there would be the need to correct incorrect factors of $2C_{F}$. This would worsen the computational efficiency. Whether the decreased efficiency is mitigated by the relative infrequency of $g\rightarrow q\bar{q}$ transitions in a typical shower is beyond the scope of this paper.

\subsection{The effects of momentum conservation}
\label{sec:momentumcons}

In the paper so far we only briefly mentioned momentum conservation, which is vital for any implementation in an event generator, and needs to be treated very carefully. Bad implementations of momentum conservation have the potential to modify the phase-space boundaries of partons in the cascade or the matrix elements, leading to NLL errors \cite{Dasgupta:2018nvj}. In a dipole shower, emissions are on-shell and their momentum is typically expressed using three components: momentum longitudinal to the emitter, momentum transverse to the emitting dipole and momentum in the `backwards' direction (collinear to the other parton in the dipole). A momentum map is used after an emission to ensure energy-momentum conservation in the shower by distributing `recoil' across the partons in an event whilst keeping the partons are on-shell.

In \cite{Forshaw:2020wrq} we presented a momentum map with the idea of being as simple as possible whilst preserving the matrix elements computed by the shower. In the map, longitudinal recoil is trivially handled correctly (it is conserved between the emission and the parent parton as dictated by the dipole partitioning) and does not spoil anything. The other components are handled by a Lorentz boost and a global re-scaling of every momentum in the event after the emission. The emission kernels are invariant under both of these (as both $z^{ab}$ and $\td k^{ab}_{\bot}/ k^{ab}_{\bot}$  are invariant under boosts and re-scalings). Thus only the phase-space is modified by the momentum map, not matrix elements. In Section 3.1 of \cite{Forshaw:2019ver} we showed that the changes to the phase-space due to recoil will generally not produce a log-enhanced term at $\mathcal{O}(\As^{2})$ and that, for global two-jet observables such as thrust, artifacts in the phase-space from the recoil after iterated emissions produce terms beyond NLL. Alternative global momentum maps with similar constructions have also been studied in \cite{Dasgupta:2020fwr} where the NLL accuracy of the maps was demonstrated for a wide range of observables. The momentum maps in \cite{Dasgupta:2020fwr} were designed so that their action preserved key features of the Lund plane \cite{Andersson:1988ee,Gustafson:1992uh} (for instance preserving the separation between emissions on the plane). They have the added benefit of conserving `backwards' components of momentum locally in a dipole, minimising the affect of the map on the phase-space available to partons in the shower. Any of these global prescriptions could be implemented into our shower without effecting the results in this paper.

\section{Errors in other dipole showers}

In this section we want to emphasize the role of the dipole partitioning to our findings. To eliminate sub-leading colour errors, the partitioning function $g_{ab}$ must satisfy
\begin{align}               
\frac{(\td\cos\theta_{aq}) \; \td \phi^{(a)}_{q}}{4\pi}  \; \int^{2\pi}_{0} \frac{\td \phi^{(a)}_{q}}{2\pi} g_{ab} \, w_{ab} = P_{ab}^{[a]} + \TT{negligible}. \label{eq;criteria}
\end{align}
In \ref{sec:limitations} we discuss the term labelled `negligible'; the remainder after azimuthal averaging when compared to the strict angular ordering result. Our dipole algorithm was carefully constructed to not produce such a contribution at all. Note that the demand of $P_{ab}^{[a]}$ being proportional to a theta function cannot be satisfied with a zero remainder if $g_{ab}$ is positive definite and only zero at a finite number of points in the phase-space. On top of this, since $P_{ab}^{[a]}$ has no dependence on the energies of the partons in the dipole, any partitioning that retains such a dependence after azimuthal averaging will result in a non-zero contribution remainder.

An interesting example to illustrate how wrong results can be obtained is that of Catani-Seymour (CS) dipole factorisation. The errors due to using a CS factorisation to construct the dipole partitioning have been previously noted in \cite{Dasgupta:2018nvj}. Here we give a complementary discussion. The CS partitioning contains both the issues described in the previous paragraph; the partitioning function is positive definite and has strong dependence on parton energies after azimuthal averaging. The partitioning that generates Catani-Seymour dipole factorisation is 
\begin{align}
g_{ab} = \frac{(k^{ab}_{\bot})^{2}\; p_{a}\cdot p_{b}}{2 p_{a} \cdot p_{q} \; (p_{a}+p_{b}) \cdot p_{q}} \equiv \frac{e^{2\eta}}{1+e^{2\eta}},
\end{align}
where $\eta$ is the dipole-frame rapidity of parton $q$ ($\eta \rightarrow \infty$ as $p_{q}/E_{q} \rightarrow p_{a}/E_{a}$ and $\eta \rightarrow -\infty$ as $p_{q}/E_{q} \rightarrow p_{b}/E_{b}$). We must compute
\begin{align}
\int^{2\pi}_{0} \frac{\td \phi^{(a)}_{q}}{2\pi} g_{ab} \, w_{ab} = \int^{2\pi}_{0} \frac{\td \phi^{(a)}_{q}}{2\pi} \frac{E_{q}^{2} \; p_{a}\cdot p_{b}}{p_{a} \cdot p_{q} \; (p_{a}+p_{b}) \cdot p_{q}}.
\end{align}
Using the basis $$p_{a} = E_{a}(1,0,0,1),$$ $$p_{b} = E_{b}(1,\sin \theta_{ab} ,0,\cos \theta_{ab}),$$ $$p_{q} = E_{q}(1,\sin \theta_{aq} \cos \phi^{(a)}_{q},\sin \theta_{aq} \sin \phi^{(a)}_{q},\cos \theta_{aq})$$
gives
\begin{align}
&\int^{2\pi}_{0} \frac{\td \phi^{(a)}_{q}}{2\pi} g_{ab} \, w_{ab} = \nonumber \\ &  \int^{2\pi}_{0} \frac{\td \phi^{(a)}_{q}}{2\pi} \frac{(1-\cos \theta_{ab})}{(1-\cos \theta_{aq}) \; (D - \sin \theta_{ab} \sin \theta_{a q} \sin \phi^{(a)}_{q})},
\end{align}
where $$ E_{b} \, D = E_{a} + E_{b}-E_{a}\cos \theta_{aq} - E_{b}\cos \theta_{ab} \cos \theta_{a q}.$$
Note $D>\sin \theta_{ab} \sin \theta_{a q}$ for all momentum configurations. It is therefore easily shown that 
\begin{align}
&\int^{2\pi}_{0} \frac{\td \phi^{(a)}_{q}}{2\pi} g_{ab} \, w_{ab} = \nonumber \\ & \qqquad \frac{(1-\cos \theta_{ab})}{(1-\cos \theta_{aq}) \; \sqrt{D^{2} - \sin^{2} \theta_{ab} \sin^{2} \theta_{a q}}}.
\end{align}
For all momentum configurations other than $\theta_{ab} = \pi$ and $E_{b}=E_{a}$ this results in 
\begin{align}
 W^{[a]}_{ab} = \frac{(\td\cos\theta_{aq}) \; \td \phi^{(a)}_{q}}{4\pi}  \; \int^{2\pi}_{0} \frac{\td \phi^{(a)}_{q}}{2\pi} g_{ab} \, w_{ab}  \not\approx P_{ab}^{[a]}.
\end{align}
Using this azimuthal averaging of the Catani-Seymour partitioning we can compute the azimuthally-averaged squared matrix element in the limit that emissions are strongly ordered in angle and energy:
\begin{align}
|\mathcal{M}_{2}|^{2} & \frac{\td^{3} \vec{p}_{q}}{2E_{q}} \approx \nonumber \\ & \frac{2\As}{\pi} \bigg[ \left(C_{\TT{F}} W^{[i]}_{ij} + \frac{C_{\TT{A}}}{2}W^{[j]}_{ji} \right)\frac{\td E_{q}}{E_{q}} \Theta(k^{ij}_{\bot} < k_{\bot}) \nonumber \\ 
&+ \left(\frac{C_{\TT{A}}}{2} W^{[j]}_{jk} + C_{\TT{F}} W^{[k]}_{kj}\right) \frac{\td E_{q}}{E_{q}} \Theta(k^{jk}_{\bot} < k_{\bot}) \bigg] \nonumber \\ & \quad \times |\mathcal{M}_{1}(\vec{p}_{i},\vec{p}_{j},\vec{p}_{k})|^{2}. 
\end{align}
As $k^{ab}_{\bot}\approx k^{ca}_{\bot} \approx  E_{q}\theta_{aq}$ in this limit and since energies are strongly ordered, the $k_{t}$ ordering theta functions are saturated and we find:
\begin{align}
|\mathcal{M}_{2}|^{2} \frac{\td^{3} \vec{p}_{q}}{2E_{q}} \approx & \frac{2\As}{\pi} \bigg[ C_{\TT{F}} W^{[i]}_{ij} + \frac{C_{\TT{A}}}{2}\left(W^{[j]}_{ji}  + W^{[j]}_{jk}\right) \nonumber \\ & + C_{\TT{F}} W^{[k]}_{kj} \bigg] \frac{\td E_{q}}{E_{q}} |\mathcal{M}_{1}(\vec{p}_{i},\vec{p}_{j},\vec{p}_{k})|^{2}. 
\end{align}
We can subtract this from the correct result (for rows 1 through 3 of Table \ref{tbl:limits}) to find the error:
\begin{align}
&\delta |\mathcal{M}_{2}|^{2} \frac{\td^{3} \vec{p}_{q}}{2E_{q}} \approx \frac{2\As C_{\TT{F}}}{\pi} \bigg[  (P^{[i]}_{ij} - W^{[i]}_{ij} + P^{[k]}_{kj}-W^{[k]}_{kj}) \nonumber \\
&+ \frac{C_{\TT{A}}}{2 C_{\TT{F}}}\left(2P^{[j]}_{ji} - W^{[j]}_{ji}  - W^{[j]}_{jk}\right)  \bigg] \frac{\td E_{q}}{E_{q}} |\mathcal{M}_{1}(\vec{p}_{i},\vec{p}_{j},\vec{p}_{k})|^{2} \nonumber \\ & \not\approx 0. \label{eq:standarderror}
\end{align}
Of course the error vanishes if $C_{\TT{F}} = C_{\TT{A}}/2$. The error becomes large when $E_{j} \ll E_{i} \approx E_{k}$. In this limit
\begin{align}
 W^{[j]}_{j b} \approx \frac{(\td\cos\theta_{jq}) \; \td \phi^{(j)}_{q}}{4\pi}  \; \frac{(1-\cos \theta_{jb})}{(1-\cos \theta_{jq}) \;  |\cos \theta_{jb} - \cos \theta_{j q}|},
\end{align}
when $\theta_{ij} \not\approx \theta_{j q}$ and $\theta_{jk} \not\approx \theta_{j q}$\footnote{In the region where $\theta_{ij} \not\approx \theta_{j q}$ terms in $E_{j}/E_{i}$ are not negligible as they screen the divergence $\theta_{ij} = \theta_{j q}$.}, and where $b = i,k$. Also in this limit 
\begin{align}
 W^{[b]}_{b j} \approx \frac{(\td\cos\theta_{iq}) \; \td \phi^{(i)}_{q}}{4\pi}  \; \frac{E_{b}(1-\cos \theta_{bj})}{E_{a}(1-\cos \theta_{bq})^{2}},
\end{align}
once again this is only valid when $\theta_{ij} \not\approx \theta_{j q}$ and $\theta_{jk} \not\approx \theta_{j q}$ (equivalently $\theta_{i q} \not\approx 0$ and $\theta_{k q} \not\approx 0$). Thus 
\begin{align}
&\delta |\mathcal{M}_{2}|^{2} \frac{\td^{3} \vec{p}_{q}}{2E_{q}} \approx \nonumber \\ & \frac{2\As C_{\TT{F}}}{\pi} \bigg[  \frac{(\td\cos\theta_{iq}) \; \td \phi^{(i)}_{q}}{4\pi} \left(\frac{\Theta(\theta_{iq} < \theta_{ij})}{1 - \cos \theta_{iq}}  - \frac{E_{b}(1-\cos \theta_{ij})}{E_{a}(1-\cos \theta_{iq})^{2}} \right) \nonumber \\ & + (i \leftrightarrow k) + \frac{C_{\TT{A}}}{2 C_{\TT{F}}}\frac{(\td\cos\theta_{jq}) \; \td \phi^{(j)}_{q}}{4\pi} \nonumber \\ & \times \bigg( 2\frac{\Theta(\theta_{jq} < \theta_{ij})}{1 - \cos \theta_{jq}} - \frac{(1-\cos \theta_{ij})}{(1-\cos \theta_{jq}) \;  |\cos \theta_{ij} - \cos \theta_{j q}|}  \nonumber \\
& - \frac{(1-\cos \theta_{jk})}{(1-\cos \theta_{jq}) \;  |\cos \theta_{jk} - \cos \theta_{j q}|}\bigg)  \bigg] \frac{\td E_{q}}{E_{q}} |\mathcal{M}_{1}(\vec{p}_{i},\vec{p}_{j},\vec{p}_{k})|^{2}. \label{eq:standarderror3}
\end{align}
Note that the $C_{\TT{A}}/C_{\TT{F}}$ piece contains a non-cancelling collinear pole when $\theta_{j q} \rightarrow 0$ and so is capable of generating logarithms in observables that probe secondary emissions even in the limit of a strong angular hierarchy, where $\theta_{j q} \ll \theta_{ij} \ll \theta_{(ij) k}$, since the numerator of $W^{[j]}_{j i}$ goes as $\mathcal{O}(\theta_{ij}^{2})$ whilst the numerator of $P^{[j]}_{j i}$ goes as $\mathcal{O}(1)$. Also note that this error cannot be fixed by using the dynamic colour factors $\mathcal{C}_{i J}(\theta_{i q},\theta_{L J})$ since in the limit we are considering the dynamic colour factors reduce exactly to the usual colour factors already present in Eq.~\eqref{eq:standarderror3}.

We can also compare the error made using a CS-partitioned dipole shower with row 4 from Table \ref{tbl:limits}. Here we find 
\begin{align}
&\delta |\mathcal{M}_{2}|^{2} \frac{\td^{3} \vec{p}_{q}}{2E_{q}} = \frac{2\As C_{\TT{F}}}{\pi} \bigg[  (P^{[(ij)]}_{(ij)k} - W^{[i]}_{ij} + P^{[k]}_{k(ij)}-W^{[k]}_{k(ij)}) \nonumber \\
&\quad - \frac{C_{\TT{A}}}{2 C_{\TT{F}}}\left( W^{[j]}_{ji}  + W^{[j]}_{jk}\right)  \bigg] \frac{\td E_{q}}{E_{q}} |\mathcal{M}_{1}(\vec{p}_{i},\vec{p}_{j},\vec{p}_{k})|^{2} \not\approx 0. \label{eq:standarderror2}
\end{align}
This error is potentially LL, since with a strong hierarchy in emission energies and angles the functions $W$ are singular and so capable of generating double logarithms. Of course this too vanishes if $C_{\TT{F}} = C_{\TT{A}}/2$. If the dipole shower instead used colour factors $\mathcal{C}_{i J}(\theta_{i q},\theta_{L J})$ this limit would be improved since the error would instead become
\begin{align}
&\delta |\mathcal{M}_{2}|^{2} \frac{\td^{3} \vec{p}_{q}}{2E_{q}} = \frac{2\As C_{\TT{F}}}{\pi} \bigg[  (P^{[(ij)]}_{(ij)k} - W^{[i]}_{ij} + P^{[k]}_{k(ij)}-W^{[k]}_{k(ij)})\nonumber \\ 
& \quad + \Theta(\theta_{j q} > \theta_{L J})(W^{[j]}_{ji}   + W^{[j]}_{jk}) - \frac{C_{\TT{A}}\Theta(\theta_{j q} < \theta_{L J})}{2 C_{\TT{F}}}\nonumber \\
&\quad \times \left( W^{[j]}_{ji}  + W^{[j]}_{jk}\right)  \bigg] \frac{\td E_{q}}{E_{q}} |\mathcal{M}_{1}(\vec{p}_{i},\vec{p}_{j},\vec{p}_{k})|^{2} \approx 0.
\end{align}
However, this improvement may not extend to higher orders since $\theta_{L J}$ as computed with the CS dipole shower branching history will not necessarily equal $\theta_{L J}$ as computed from a branching history matched to the angular-ordered description. This problem, combined with Eq.~\eqref{eq:standarderror}, is sufficient for us to assert that CS dipole showers employing the dynamic colour factors $\mathcal{C}_{i J}(\theta_{i q},\theta_{L J})$ will still be subject to LL errors in some observables that angular-ordering can completely describe at LL.

\section{Conclusions}
We have performed a fixed-order cross-check of the dipole shower presented in  \cite{Forshaw:2020wrq} and shown that the shower performs as it was designed to: the shower inherits its handling of collinear radiation from an angular-ordered shower whilst improving over angular ordering in the case of the leading colour, wide-angle soft radiation. We also highlight a limitation of our original approach, showing how the dipole shower will not assign correct colour factors to emissions disordered in angle, though they will be correct at leading colour. We then introduced a new method for correcting these colour factors. The new method is efficient: the computation time on average grows logarithmically with parton multiplicity. Using this method, our shower will match the LL accuracy of an angular-ordered shower in cases where an angular-ordered shower has LL accuracy. When enhanced with the CMW running coupling \cite{Catani:1996vz}, our shower will include all leading logarithms and leading-colour, next-to-leading logarithms in the two-jet limit for \newline continuously-global observables. As it stands, the shower will not be capable of the full-colour NLL resummation of global observables, due to the absence of full colour $g \rightarrow q \bar{q}$ transitions. These transitions could be included as described in \cite{Forshaw:2020wrq} but would generate sub-leading colour NLL errors: however, they could be included at full colour by extending the methods outlined in Section \ref{sec:correcting}.

\section*{Acknowledgements}
This work has received
funding from the UK Science and Technology Facilities Council (grant
no. ST/P000800/1), the European Union’s Horizon 2020 research and
innovation programme as part of the Marie Skłodowska-Curie Innovative
Training Network MCnetITN3 (grant agreement no. 722104), and in part
by the by the COST actions CA16201 ``PARTICLEFACE'' and CA16108
``VBSCAN''. JH thanks the UK Science and Technology Facilities Council
for the award of a studentship. We would also like to thank the organisers of the ``Taming the accuracy of event generators'' workshop (2020) for facilitating enlightening discussions. We especially want to thank Gavin Salam for valuable discussions. 

\section*{Additional note}

As this paper was being concluded, reference \cite{Hamilton:2020rcu} appeared. One solution that \cite{Hamilton:2020rcu} provides to improve dipole shower colour by dividing the emission phase-space and subsequently assigning colour factors $C_{\TT{F}}$ or $C_{\TT{A}}/2$ in accordance with QCD coherence, appears to be similar to the solution presented here. The authors of \cite{Hamilton:2020rcu} state that their method for correctly assigning colour factors ``... can be applied to almost any dipole or antenna shower.''

\appendix

\section{Drawing planar diagrams at arbitrary order}
\label{sec:diagrams}

In this appendix we demonstrate that the planar diagrams representing re-arrangements of our dipole shower into an angular ordered shower, in Figure \ref{fig:dipoleerrordiagram}, are not just a feature of our dipole shower at $\mathcal{O}(\As^2)$ but rather can continue to be used at higher orders if we continue to assume that the branching history produced by our shower has a strong hierarchy in angles. We do not assume a hierarchy in angles that is concurrently ordered with their $k_{t}$. At a scale $k_{\bot}$, a given $n$-parton state produced by our dipole shower has a weight at a point in the $n$-parton phase-space $\td \mathcal{S}_{n}(k_{\bot})$. We consider dressing this state with one further gluon, $q$, produced by the shower. This gives an $(n+1)$-parton state:
\begin{align}
&\td \mathcal{S}_{n+1}(k_{\bot}) \,\frac{\td^{3} \vec{p}_{q}}{2E_{q}} = \nonumber \\
& \quad \frac{\As}{\pi} \sum_{a,b \; \TT{c.c.}} \bigg(\mathcal{C}_{a J}(\theta_{a q},\theta_{L J}) \, g_{ab} \, \mathcal{P}^{\TT{d}}_{a \rightarrow a q}(z^{ab}) \td z^{ab} \nonumber \\
& \qquad + (a \leftrightarrow b)\bigg) \frac{\td \ln k^{ab}_{\bot} \td\phi}{2\pi} \, \delta(k^{ab}_{\bot} - k_{\bot}) \; \td \mathcal{S}_{n}(k_{\bot}),
\end{align}
where $\TT{c.c.}$ means $a$ and $b$ are colour connected in the $n$-parton state and $J$ is the hard parton that initiated $a$'s branch. All other symbols have the same definitions as in the previous sections.
We can azimuthally average exactly as in Section \ref{sec:3}, and find
\begin{align}
&\td \mathcal{S}_{n+1}(k_{\bot}) \,\frac{\td^{3} \vec{p}_{q}}{2E_{q}} = \nonumber \\ 
&\qquad \frac{\As}{\pi} \sum_{a,b \; \TT{c.c.}} \bigg(\mathcal{C}_{a J}(\theta_{a q},\theta_{L J})  \, P^{[a]}_{ab} \, \mathcal{P}^{\TT{d}}_{a \rightarrow a q}(z^{ab}) \td z^{ab} \nonumber \\
& \qqquad + (a \leftrightarrow b)\bigg) \delta(k^{ab}_{\bot} - k_{\bot}) \;\td \mathcal{S}_{n}(k_{\bot}). \label{eq:4.6}
\end{align}
Just as we have already demonstrated at $\mathcal{O}(\As^{2})$, the weight assigned to the $(n+1)$-state after azimuthal averaging uses the same LC emission kernels as an angular-ordered approach. We can make this very explicit by exchanging the sum over colour lines with a sum over parton indices. To illustrate this, at LC we find
\begin{align}
&\td \mathcal{S}_{n+1}(k_{\bot}) \,\frac{\td^{3} \vec{p}_{q}}{2E_{q}} = \nonumber \\ 
&\frac{\As}{\pi} \sum_{a} \bigg(\frac{C_{\TT{A}}}{2}  \, P^{[a]}_{ab} \, \mathcal{P}^{\TT{d}}_{a \rightarrow a q}(z^{ab}) \td z^{ab} \delta(k^{ab}_{\bot} - k_{\bot}) \nonumber \\
& \quad + \delta^{(g)}_{a} \frac{C_{\TT{A}}}{2}  \, P^{[a]}_{ac} \, \mathcal{P}^{\TT{d}}_{a \rightarrow a q}(z^{ac}) \td z^{ac} \delta(k^{ac}_{\bot} - k_{\bot}) \bigg)  \;\td \mathcal{S}_{n}(k_{\bot}). \label{eq:4.7}
\end{align}
We can exchange the non-singular dependence on $b$ and $c$ in Eqs.~\eqref{eq:4.6} and \eqref{eq:4.7} with that of $J$ (or the other hard parton $J'$ if either $b$ or $c$ are in the opposing hemisphere). Similarly, for the non-singular region $\theta_{aq} \gg \theta_{aJ}$ we can exchange the dependence on $a$ with $J$ so that $\theta_{aq} \approx \theta_{Jq}$. Thus, just as in an angular-ordered framework, emissions are generated with a weight $$C  \, P^{[a]}_{aJ} \, \mathcal{P}^{\TT{d}}_{a \rightarrow a q}(z^{aJ})$$ when they can probe the jet and $$C  \, P^{[J]}_{JJ'} \, \mathcal{P}^{\TT{d}}_{J \rightarrow J q}(z^{JJ'})$$ when they cannot. At LC, $C = C_{\TT{A}}/2$ when $a$ is a quark and, when $a$ is a gluon, $C = C_{\TT{A}}$ if $q$ can probe the jet (determined by the angular ordering constraint embedded in $P^{[a]}_{aJ}$\footnote{This constraint is saturated by using an angular ordering variable in an angular-ordered shower and so would typically be omitted if one where to write Eq.~\eqref{eq:4.6} specifically for such a shower.}) and $C = C_{\TT{A}}/2$ when $q$ cannot. It is these properties that our planar diagrams are defined to encapsulate, validating their usage at arbitrary higher orders. The planar diagrams led us to define $\mathcal{C}_{a J}(\theta_{a q},\theta_{L J})$ so that the sub-leading $\Nc$ terms are included in accordance with Figure \ref{fig:correctfactorsAO}.

\section{Current limitations of our dipole shower}
\label{sec:limitations}

\begin{figure*}[h]
	\centering
	\subfigure[$\theta_{ab} = 2$ and $\phi^{(a)}_{q} = 2$, i.e. $q$ is out of the plane of the dipole $(a,b)$. The choice of $\phi^{(a)}_{q} = 2$ is indicative of the majority of the emission phase-space for which  $\phi^{(a)}_{q} \not\approx 0$. ]{\includegraphics[width=0.47\textwidth]{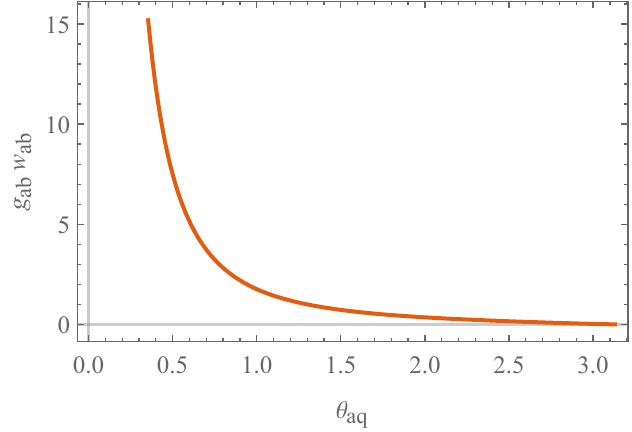}} ~~
	\subfigure[$\theta_{ab} = 2$ and $\phi^{(a)}_{q} = 0.05$, i.e. $q$ is only slightly out of the plane of the dipole $(a,b)$. Negative weights are starting to emerge, as are the peaks that become an integrable divergence for $\phi^{(a)}_{q} = 0$. ]{\includegraphics[width=0.47\textwidth]{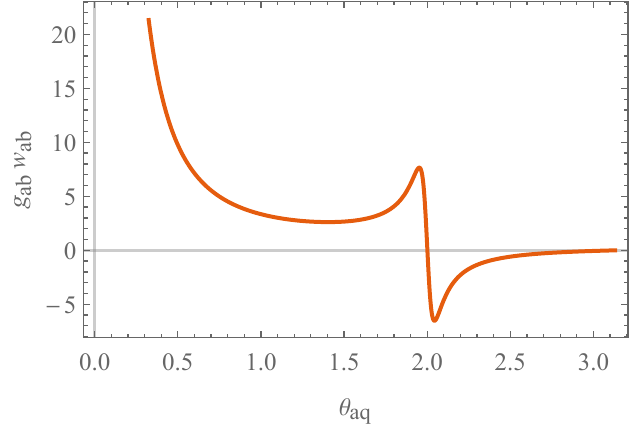}} ~~
	\subfigure[$\theta_{ab} = 2$ and $\phi^{(a)}_{q} = 0$, i.e. $q$ is in the plane of the dipole $(a,b)$. For $\theta_{aq}<\theta_{ab}$, $q$ is `inside' the dipole. The negative weights and integrable divergence are present when $\theta_{aq} = \theta_{ab}$.]{\includegraphics[width=0.47\textwidth]{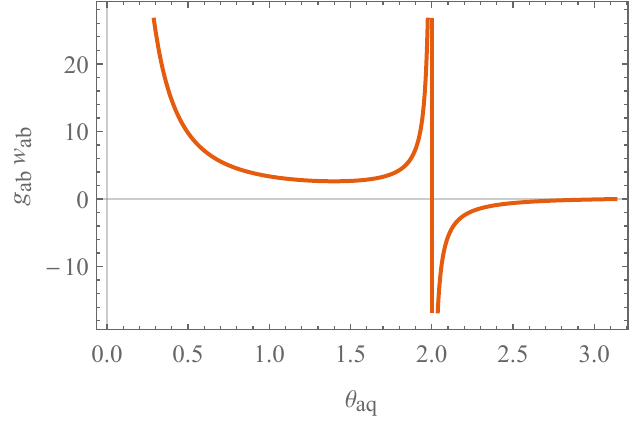}}
	\caption{Graphs of $g_{ab} w_{ab}$ as a function of $\theta_{ab}, \theta_{aq}, \phi^{(a)}_{q}$ as measured in the lab frame.}
	\label{fig:undesirable}
\end{figure*}

An important part of our dipole shower is its partitioning. However, the form of our partitioning, $g_{ab}$ (defined through Eqs.~\eqref{eq:partitioning1} and \eqref{eq:partitioning2}), might cause complications in a computational implementation of our shower. In this appendix we will discuss the issues and possible solutions.

A dipole shower is fully differential in the parton phase-space and so emits by sampling from the distribution $g_{ab} w_{ab}$ to populate a 3-dimensional phase-space for each parton.\footnote{Including hard-collinear physics, the shower samples from $g_{ab}\mathcal{P}^{\TT{d}}_{ab}$ where $\mathcal{P}^{\TT{d}}_{ab}$ is a dipole splitting function but this does not effect our discussion here.} However, $g_{ab} w_{ab}$ has two undesirable properties (illustrated in Figure \ref{fig:undesirable}): firstly $g_{ab} w_{ab}$ is negative in some portions of the emission phase-space, introducing negative weights into the shower; secondly $g_{ab} w_{ab}$ contains an integrable  singularity when $\theta_{aq} = \theta_{ab} < \pi$ and $\phi^{(a)}_{q} = 0$ (i.e. $q$ is in the plane of partons $a$ and $b$). Both of these features can be handled in a modern dipole shower: the Herwig dipole shower already contains all the necessary machinery \cite{Herwig_dipole_shower}, as do others \cite{DIRE}. However, both features will hinder numerical convergence. Fortunately the two features counter balance each other: $g_{ab} w_{ab}$ is most negative when $\theta_{aq} = (1+\varepsilon)\theta_{ab} $, for $\varepsilon \ll 1$ whilst strictly positive, and $\phi^{(a)}_{q} = 0$. The negative weights and integrable singularity are linked such that, when $\theta_{aq} = \theta_{ab}$, $g_{ab} w_{ab}$ azimuthally averages to a well behaved quantity,
\begin{align}
\; \left. \int^{2\pi}_{0} \frac{\td \phi^{(a)}_{q}}{2\pi} g_{ab} w_{ab} \right|_{\theta_{aq} = \theta_{ab}} = \frac{1}{2(1 - \cos \theta_{aq})}. \label{eq:azav}
\end{align}
A simple solution to the two issues would be, in regions bounded by $\theta_{aq} = \theta_{ab} \pm \delta\theta$ (for $\delta\theta / \theta_{ab} \ll 1$), to sample emissions according to the azimuthally averaged distribution, Eq.~\eqref{eq:azav}. This would entail sampling emissions from a discontinuous distribution but would alleviate the undesirable features whilst only introducing a power correction in $\delta\theta$ to azimuthal correlations in the shower.

Alternatively, one might use an alternative partitioning, $\tilde{g}_{ab}$, free from negative weights and integrable singularities, that satisfies
\begin{align}
\frac{(\td\cos\theta_{aq}) \; \td \phi^{(a)}_{q}}{4\pi}  \; \int^{2\pi}_{0} \frac{\td \phi^{(a)}_{q}}{2\pi} \tilde{g}_{ab} \, w_{ab} \approx P_{ab}^{[a]}. \label{eq:approxrequirement}
\end{align}
This $\tilde{g}_{ab}$ would be suitable for use with our proposed dynamic colour factors and retain our shower's accuracy concerning LC NLL physics. It is possible that a pre-existing partitioning employed by another parton shower might already achieve this. We have demonstrated that the Catani-Seymour partitioning \cite{Catani:1996vz} does not satisfy this requirement but there are others on the market that we have not tested \cite{Nagy:2008eq,Lonnblad:1992tz,DIRE,Dasgupta:2020fwr}. An acceptable partitioning should at least satisfy
\begin{align}
&\frac{(\td\cos\theta_{aq}) \; \td \phi^{(a)}_{q}}{4\pi}  \; \int^{2\pi}_{0} \frac{\td \phi^{(a)}_{q}}{2\pi} (g_{ab}-\tilde{g}_{ab}) \, w_{ab} \nonumber \\ 
&= \frac{(\td\cos\theta_{aq}) \; \td \phi^{(a)}_{q}}{4\pi(1 - \cos \theta_{aq})}\Theta(\theta_{aq} < \theta_{ab})f\left(E_{q} / E_{a}, \theta_{aq} , \theta_{ab}; ... \right),
\end{align}
where the ellipses denote all other kinematic quantities on which $f$ depends but $q$'s emission kernel otherwise does not, and where
\begin{align}
     &\left[ \int^{Q}_{\tau Q} \ln^{n} \frac{E_{a}}{Q} \;  \frac{\td E_{a}}{E_{a}} \int^{Q}_{E_{a}} \frac{\td E_{q}}{E_{q}} + \int^{Q}_{\tau Q} \frac{\td E_{q}}{E_{q}} \int^{Q}_{E_{q}} \ln^{2n-2} \frac{E_{a}}{Q} \;  \frac{\td E_{a}}{E_{a}} \right]  \nonumber \\ &\int^{1}_{\tau } \frac{\td \theta_{aq}}{\theta_{aq}} \; f\left( E_{q} / E_{a},  \theta_{aq} , \theta_{ab}; ... \right) \Theta(\nu_{q} < \nu_{a}) = \mathcal{O}(\ln^{n+1} \tau),\label{eq:allowablediffs1}
\end{align}
\begin{align}
     &\left[ \int^{1}_{\tau} \ln^{n} \theta_{ab}  \; \frac{\td \theta_{ab}}{\theta_{ab}} \int^{1}_{\theta_{ab}} \frac{\td \theta_{aq}}{\theta_{q}} + \int^{1}_{\tau} \frac{\td \theta_{q}}{\theta_{aq}} \int^{1}_{\theta_{aq}} \ln^{2n-2} \theta_{ab}  \; \frac{\td \theta_{ab}}{\theta_{ab}} \right] \nonumber \\ &
     \times \int^{Q}_{\tau Q} \frac{\td E_{q}}{E_{q}}\; f\left( E_{q} / E_{a},  \theta_{aq} , \theta_{ab}; ... \right) \Theta(\nu_{q} < \nu_{a})  = \mathcal{O}(\ln^{n+1} \tau),\label{eq:allowablediffs2}
\end{align}
where $\nu_{q,a}$ is the shower ordering variable. These ensure that $f$ at most contributes logarithms of the form $\As^{n} L^{2n-2}$ to the expansion of an observable. In most two-jet event shape observables, towers of $\As^{n} L^{2n-2}$ logarithms which first appear for $n = 1$ are NNLLs in the resummed observable. If one were to perform these tests using the Catani-Seymour partitioning, each of Eqs.~\eqref{eq:allowablediffs1} and \eqref{eq:allowablediffs2} evaluates to $\mathcal{O}(\ln^{n+3} \tau)$; a LL error (the calculation of which follows almost exactly the same structure as the thrust calculation in \cite{Dasgupta:2018nvj}).

\bibliographystyle{spphys}
\bibliography{crosschecks}

\end{document}